\documentclass[sigconf,nonacm,natbib=false]{acmart}

\usepackage{pifont}
\usepackage{xcolor}

\usepackage{tikz}
\usepackage{amsmath}
\usepackage{booktabs} 
\usepackage{multirow}
\usepackage{color,soul}
\usepackage{algorithm,float}
\usepackage[noend]{algpseudocode}
\usepackage{graphicx}
\usepackage{subfig}
\usepackage{comment}
\usepackage{enumitem,kantlipsum}
\usepackage{textcase}
\usepackage{makecell}
\usepackage{longtable}
\usepackage{multirow}
\usepackage{tabularx}
\usepackage{array}
\usepackage{balance}
\usepackage{url}

\newcommand{\cmark}{\ding{51}} 
\newcommand{\xmark}{\ding{55}} 
\newcommand{\pmark}{\ding{117}} 

\algrenewcommand\algorithmiccomment[1]{\hfill$\triangleright$ #1}

\def\BibTeX{{\rm B\kern-.05em{\sc i\kern-.025em b}\kern-.08em
    T\kern-.1667em\lower.7ex\hbox{E}\kern-.125emX}}
\begin{document}

\title{RINSER: Accurate API Prediction Using Masked Language Models}

\author{Muhammad Ejaz Ahmed}
\affiliation{%
  \institution{CSIRO Data61}
  \country{Australia}
}
\email{ejaz.ahmed@data61.csiro.au}

\author{Christopher Cody}
\affiliation{%
  \institution{CSIRO Data61}
  \country{Australia}
}
\email{christopher.cody@data61.csiro.au}

\author{Muhammad Ikram}
\affiliation{%
  \institution{Macquarie University}
  \country{Australia}
}
\email{muhammad.ikram@mq.edu.au}

\author{Sean Lamont}
\affiliation{%
  \institution{DSTG}
  \country{Australia}
}
\email{sean.lamont2@defence.gov.au}

\author{Alsharif Abuadbba}
\affiliation{%
  \institution{CSIRO Data61}
  \country{Australia}
}
\email{sharif.abuadbba@data61.csiro.au}

\author{Seyit Camtepe}
\affiliation{%
  \institution{CSIRO Data61}
  \country{Australia}
}
\email{seyit.camtepe@data61.csiro.au}

\author{Surya Nepal}
\affiliation{%
  \institution{CSIRO Data61}
  \country{Australia}
}
\email{surya.nepal@data61.csiro.au}

\author{Muhammad Ali Kaafar}
\affiliation{%
  \institution{Macquarie University}
  \country{Australia}
}
\email{dali.kaafar@mq.edu.au}

\begin{abstract}
Malware authors commonly use obfuscation to hide API identities in binary files, making analysis difficult and time-consuming for a human expert to understand the behavior and intent of the program. Automatic API prediction tools are necessary to efficiently analyze {\it unknown} binaries, facilitating rapid malware triage while reducing the workload on human analysts. In this paper, we present RINSER (Accu\textbf{\underline{R}}ate AP\textbf{\underline{I}} predictio\textbf{\underline{N}} using ma\textbf{\underline{S}}ked languag\textbf{\underline{E}} model lea\textbf{\underline{R}}ning), an automated framework for predicting Windows API (WinAPI) function names. RINSER introduces the novel concept of API codeprints, a set of API-relevant assembly instructions, and supports x86 PE binaries. RINSER relies on BERT's masked language model (LM) to predict API names at scale, achieving 85.77\% accuracy for normal binaries and 82.88\% accuracy for stripped binaries. We evaluate RINSER on a large dataset of 4.7M API codeprints from 11,098 malware binaries, covering 4,123 unique Windows APIs, making it the largest publicly available dataset of this type. RINSER successfully discovered 65 obfuscated Windows APIs related to C2 communication, spying, and evasion in our dataset, which the commercial disassembler IDA failed to identify~\cite{idapro}. 
Furthermore, we compared RINSER against three state-of-the-art approaches, showing over 20\% higher prediction accuracy. We also demonstrated RINSER's resilience to adversarial attacks, including instruction randomization and code displacement, with a performance drop of no more than 3\%.  
\end{abstract}
\maketitle

\section{Introduction} \label{sec:introduction}
Predicting the names of Windows API functions in binaries can be highly useful for many security applications including
malware analysis~\cite{downing2021deepreflect, xu2021hawkeye,cheng2018towards,jin2022symlm,he2018debin},
capability analysis~\cite{alrawi2021forecasting, downing2021deepreflect, capa}, 
vulnerability identification~\cite{lingmalgraph, cochard2022investigating, downing2021deepreflect},
reverse engineering~\cite{park2021identifying, pucher2022identification, aghakhani2020malware},
and compatibility testing~\cite{liu2014research, sawant2012software}.
Unfortunately, malware authors may attempt to complicate the analysis by obfuscating the names of Windows API functions~\cite{suenaga2009museum, mantovani2020prevalence, cheng2021obfuscation} in a binary to make it difficult for security analysts and anti-malware scanners to analyse and detect the malware. 
In 2021, Mandiant observed that 51.4\% of analysed malicious intrusions employed obfuscation as a technique to evade detection and subsequent analysis~\cite{mandiant2022m}. 
Obfuscation techniques including the deletion or modification of Import Address Table (IAT)~\cite{cheng2021obfuscation}, API function hashing~\cite{api_hashing}, string encryption~\cite{dong2018understanding}, and API name mangling~\cite{vinciguerra2003experimentation} are commonly employed to hide API names to hinder analyses.

\begin{figure}[ht]
  \centering
  \includegraphics[width=2.9in,clip]{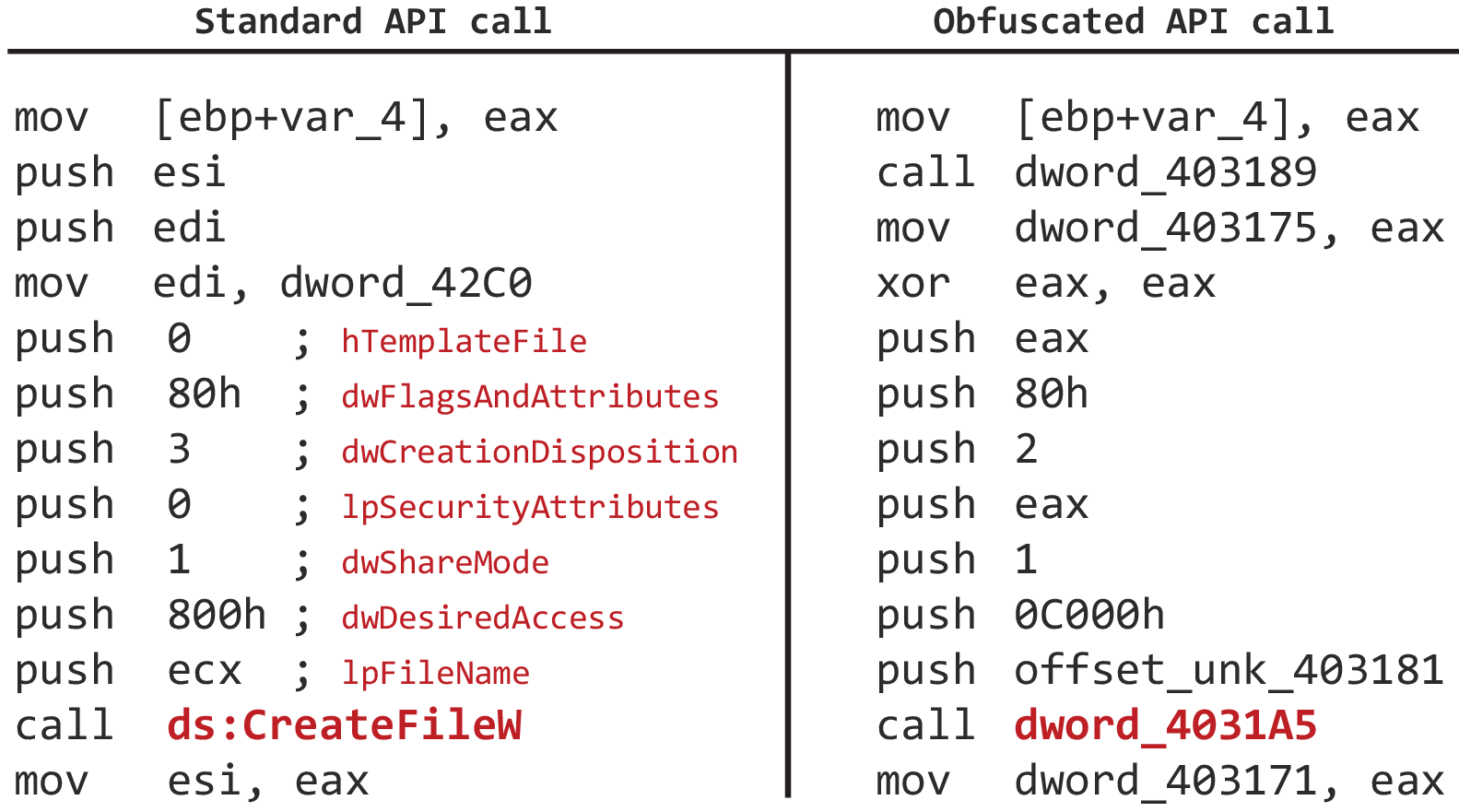}
  \caption{Real-world example illustrates the standard and obfuscated disassembly for the {\it CreateFileW} API.
  }
  \label{fig:example}
\end{figure}

API obfuscation may replace the API names with placeholders, typically dummy or meaningless words, as demonstrated in Figure~\ref{fig:example}, which are either determined at runtime or through reverse engineering the obfuscation technique to restore the API names. Human analysts can often infer API names by examining the arguments passed to them and the context in which they are invoked. For instance, as illustrated in Figure~\ref{fig:example}, the parameters {\it dwDesiredAccess}, {\it dwShareMode}, {\it dwCreationDisposition}, and {\it dwFlagsAndAttributes} are the pre-defined constants~\cite{createfileapi} indicating access control parameters, file sharing restrictions and attributes, and permission flags, respectively. They can take only a small set of potential values. For instance, the {\it dwCreationDisposition} and {\it dwShareMode} parameters only accept five and four pre-defined constant values, respectively, ranging from 1 to 5 and 1 to 4.
A human analyst observing these parameters on the stack can conclude that the {\it CreateFileW} API is called, even if the API name is not visible. However, this process is painstakingly slow and requires much manual effort and domain knowledge to infer correct API names. 


Existing approaches for predicting API names primarily focus on Linux operating system~\cite{jin2022symlm, patrick2023xfl, li2021palmtree, xi2013api, choi2015api}, with limited attention given to this problem~\cite{kotov2018towards, sharif2008eureka} for Windows OS due to obvious reasons, including the unavailability of the source. 
Kotov et al.~\cite{kotov2018towards} proposed an HMM-based method for predicting API functions' names for Windows OS based on their input parameter values, focusing on 25 popular API functions. 
However, the Windows OS comprises a large number of API functions bundled across numerous DLLs that are built into the Windows OS.
Our analysis of 11,098 PE binaries revealed the use of 4,123 unique Windows APIs, with the number of input parameters for these APIs ranging from 0 to 14.
Given the large diversity in API calls and corresponding input parameters, no existing method addresses this problem for the Windows OS.
There exists a separate stream of related research~\cite{jin2022symlm, he2018debin, ding2019asm2vec, theregister_malware} that focuses on predicting debug information, including function names, for Linux OS by training ML models on popular open-source projects.
However, obtaining source code, particularly for Windows malware, can be difficult. Only 14\% of newly identified malware families have publicly available source code, with the remaining 86\% being inaccessible~\cite{mandiant2022m}. 
Their results, even on simple, non-obfuscated binaries 
are limited and inadequate for binaries that employ sophisticated obfuscation methods.

In this paper, we present a fast and accurate approach for predicting Windows APIs (WinAPIs), called RINSER.  
RINSER is an automated WinAPI prediction framework designed for PE binaries in the x86 architecture. It learns APIs and their contextual usage from normal binaries and leverages this knowledge to accurately predict APIs in stripped (binaries with symbol/debug info, including function names, are removed) and obfuscated (binaries where APIs are obscured to avoid detection) binaries.
RINSER relies on a novel technique based on program analysis to extract the context of an API function by identifying its input parameters and the corresponding assembly instructions that are semantically related to the input parameter, known as \textit{API codeprint}. 
An API codeprint consists of the API name, names, and corresponding values of its input parameters, and a set of semantically-related assembly instructions for each input parameter value.
RINSER utilizes a masked language model (LM) to learn API codeprints for Windows API functions in an unsupervised manner. The LM task involves predicting values of randomly masked tokens within tokenized API codeprints, compelling the model to infer missing tokens based on contextual cues from adjacent tokens.
Specifically, we ask the model to infer the names of APIs for a given API codeprint by masking-out the actual name of the API.
We show that this context-aware understanding of the problem domain improves not only the model’s accuracy on regular binaries but also on stripped and obfuscated binaries with minimal fine-tuning. 
Another benefit of masked LM is that the masked token prediction task is self-supervised, and so does not require any labeled data. Thus, RINSER can be further improved for other downstream tasks, such as API prediction for stripped binaries, API parameters' values prediction, or API call sequence prediction. 
The main contributions of our study are: 
\begin{itemize}[nolistsep,leftmargin=8pt]
    \item \textit{API codeprints generation:} We propose a novel program analysis-based technique to build context for APIs\footnote{Throughout the paper, APIs refer to WinAPIs.} using a two-step learning approach: we first identify and extract assembly instructions related to the API calls and corresponding parameters (\S~\ref{ssec: api name and parameters extraction}), then we backtrack through the program disassembly based on the identified parameters to extract semantically-related assembly instructions to construct \textit{API codeprints} (\S~\ref{ssec:context extraction api parameters}). This enables RINSER to monitor the flow of information through the program, providing a broader context and avoiding the potential loss of crucial information. We show that the use of context-based API codeprints enhances API prediction performance, increasing it from 20\% to 60\% when compared to scenarios where codeprints are not used (see ablation study in \S~\ref{ssec: ablation study}).
    \item \textit{Building API prediction model:} We demonstrate that the masked Language Model on API codeprints is an effective pretraining task for predicting the names of obfuscated APIs as it forces the model to learn the semantics of the API codeprints. We then show how the learned model can be leveraged with fine-tuning to accurately predict the names of APIs for normal, stripped, and obfuscated binaries (\S~\ref{ssec: masked lm pretraining} and \ref{ssec:finetuning}). We compared RINSER against three state-of-the-art approaches and found that it significantly outperforms existing solutions (\S~\ref{ssec: comparison and related work}). Furthermore, we demonstrate RINSER's robustness against adversarially-crafted binaries, with minimal impact on its performance (\S~\ref{sec: robustness}).
    \item \textit{Curation and release of a large and diverse API dataset:} We constructed a large ground truth dataset using IDA's~\cite{idapro} annotations/comments feature to extract 4,744,969 API codeprints from 11,098 malware binaries, covering 4,123 unique WinAPIs -- the largest WinAPIs dataset publicly available. In addition, we construct a reference database of Windows API names and their input parameters, as observed in malware binaries in the wild. 
    \item \textit{Evaluation against diverse sets of binaries:} RINSER achieved an overall 85.77\% and 82.88\% accuracy in predicting the names of APIs for normal and stripped binaries, respectively. In our dataset, RINSER discovered 148,685 instances where API names were missing, i.e., unresolved by IDA, potentially due to obfuscation. RINSER was able to correctly predict the names of 69,921 of these obfuscated APIs from 2,918 binaries, where IDA failed to identify the names of these APIs.
    Furthermore, we provided a detailed ablation study and model explanation, along with a macro-average analysis of APIs (\ref{ssec: ablation study}). 
    
\end{itemize}

\section{Background and Threat Model} \label{sec : background}
\subsection{Standard calling convention for functions}
We begin by presenting an overview of how Windows API functions are invoked and how their parameters are passed. The calling convention of a function defines how the parameters are passed to the function, whether in CPU registers, on the stack, or using a combination of both, and who is responsible for cleaning up the stack after the function returns~\cite{eagle2011ida}
On 32-bit (x86) systems, many Win32 API functions use the {\it stdcall} convention, where parameters are passed on the stack in right-to-left order, but the {\it callee} cleans up the stack after execution.
The reason for placing the parameters on the stack in the right-to-left order is that the leftmost (first) parameter of the function will always be at the top of the stack when the function is called~\cite{eagle2011ida,idapro}. 
Passing arguments in right-to-left order ensures that the first (leftmost) parameter ends up deeper in the stack, making it easier to support functions with a variable number of arguments, where the number of parameters isn't known in advance.
On x86 architectures, function parameters are typically passed on the stack in right-to-left order, as required by calling conventions like {\it stdcall} and {\it cdecl}. While early compilers often used ``push'' instructions for each argument, modern compilers usually use ``mov'' or ``lea'' instructions to store arguments directly into the stack frame, as this allows for better optimization and alignment before invoking the function with a call instruction~\cite{eagle2011ida}.
\begin{figure}[h]
  \centering
  \includegraphics[width=3.2in,clip]{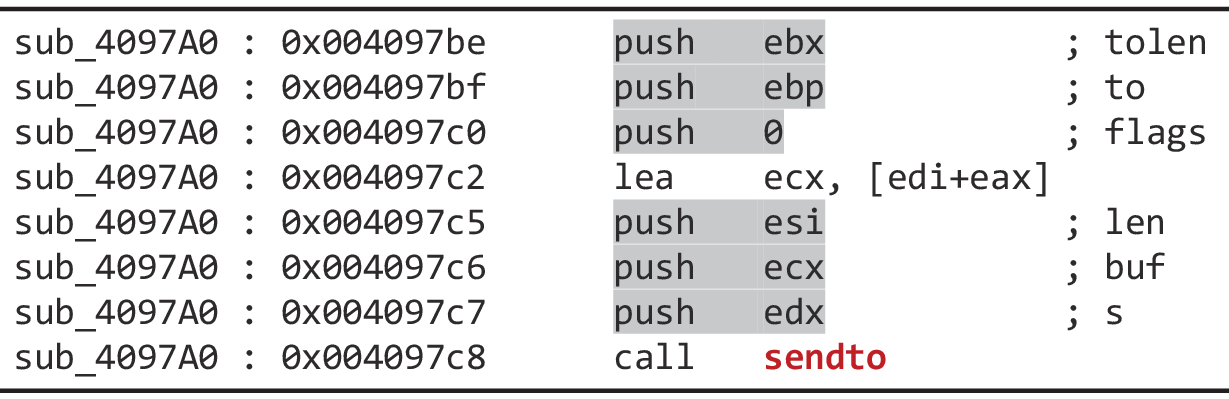}
  \caption{A real-world example of the {\it sendto} API call with its parameters pushed onto the stack (highlighted). Comments (';') show parameter names generated by IDA.}
  \label{fig:apicall}
\end{figure}

Figure~\ref{fig:apicall} illustrates the assembly instructions for a malware binary file disassembled using IDA.
API \textit{sendto} (see Microsoft SDK documentation~\cite{ms_sdk}) expecting six parameters to be pushed in right-to-left order.
The API parameters are either placed in CPU registers (e.g., {\it tolen}, {\it to}, {\it len}, {\it buf}, {\it s}) or pushed directly to the stack when parameters are pre-defined constant (e.g., {\it flags}). Also, note that parameters are pushed to the stack in right-to-left order.
The names of the API parameters are provided by IDA FLIRT (fast library identification and recognition technology)~\cite{ida_flirt_tech}.
IDA FLIRT can extract the names of API parameters and other debug information from a database of known standard API function signatures and annotate them in the program being disassembled. For example, the input parameter names are represented after `;', where the parameter values that are passed to the API are represented by the operands of {\it push} instructions, as shown in Figure~\ref{fig:apicall}.
Human experts rely on the annotations generated by IDA to analyse binaries to identify their intents. 

Table~\ref{tab:api-visibility} shows API calls' visibility across normal, stripped, and obfuscated binaries. It highlights the increasing complexity of reverse engineering as binaries transition from normal forms (relatively easy and less time-consuming) to obfuscated forms (much harder and time-consuming).

\begin{table}[h!]
\centering
\renewcommand{\arraystretch}{1.2}
\begin{tabular}{>{\raggedright}p{1.5cm} p{1.8cm} p{1.5cm} p{2cm}}
\toprule
\textbf{Aspect} & \textbf{Normal Binary} & \textbf{Stripped Binary} & \textbf{Obfuscated Binary} \\
\midrule
API Function Names (e.g., \texttt{CreateFileA}) 
& \cmark\ Visible in IAT and disassembly 
& \pmark\ Partially visible; missing from symbols 
& \xmark\ Obfuscated via hashing/encryption or resolved via \texttt{GetProcAddress} \\
\addlinespace
Import Address Table (IAT) 
& \cmark\ Full and usable 
& \cmark\ Present 
& \xmark\ Often bypassed via runtime resolution \\
\addlinespace
API Call Resolution 
& \cmark\ Direct (e.g., \texttt{call CreateFileA}) 
& \pmark\ Mostly direct 
& \xmark\ Indirect (e.g., via runtime decryption) \\
\addlinespace
Reverse Engineering Difficulty 
& Low (explicit API use) 
& Medium (missing names) 
& High (encrypted/indirect APIs) \\
\bottomrule
\end{tabular}
\caption{Comparison of API visibility and analysis difficulty across normal, stripped, and obfuscated PE binaries.}
\label{tab:api-visibility}
\end{table}

\subsection{Threat Model}
We consider an attacker who employs techniques to obfuscate APIs~\cite{suenaga2009museum} or remove debug information~\cite{jin2022symlm, he2018debin} before compiling source code into Windows PE binaries. 
Such tactics aim to obscure the code and thwart analysis by security analysts and anti-malware scanners. The obfuscation of APIs and binary stripping create challenges in identifying the true behavior and intent of the binaries. We focus our study on Windows PE binaries compiled for x86 architecture to uncover concealed threats and hidden capabilities resulting from these obfuscation and stripping methods. By examining the given Windows PE binaries, we aim to reveal the intended purpose of the code based on the use of obfuscated APIs, without access to the original source code. Our goal is to identify and understand the malicious behaviors or hidden functions within the binaries despite the lack of source code transparency.

\section{Design of RINSER} \label{sec : design}


\begin{figure*}[!ht]
  \centering
  \includegraphics[width=\linewidth,clip]{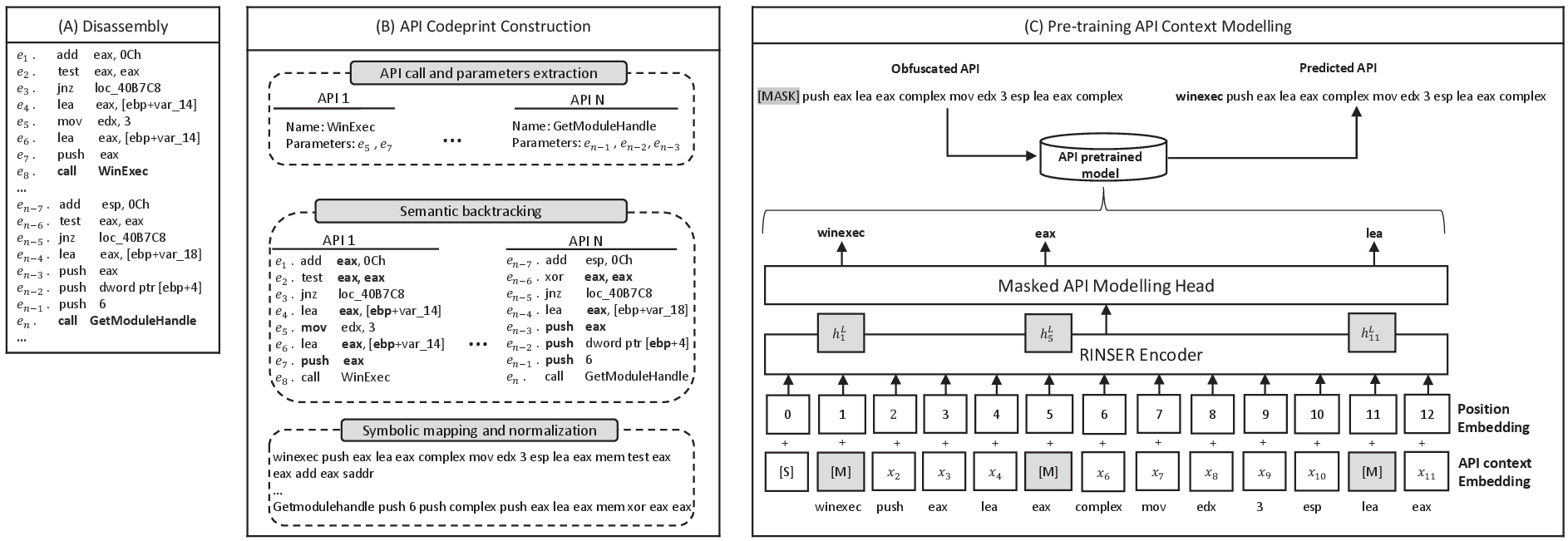}
  \caption{\small 
  An overview of RINSER. 
  (A) We first disassemble the input binary using IDA Pro. (B) Then, we build API context by: (1) identifying API callsites and input parameters from disassembly (\S~\ref{ssec: api name and parameters extraction}), (2) backtracking, starting from parameters, in the disassembly to extract semantically-related assembly instructions for each API, resulting in API codeprints (\S~\ref{ssec:context extraction api parameters}). (3) we perform symbolic mapping and normalization of API codeprints and input them to the training step (\S~\ref{ssec:symbollic mapping}).
  (C) We then pretrain the model with the masked LM task using API codeprints to perform automatic API deobfuscation for normal binaries (\S~\ref{ssec: masked lm pretraining}), and then we finetune the pretrained model to deobfuscate APIs for stripped binaries  \S~\ref{ssec:finetuning}).}
  \label{fig:system}
\end{figure*}

\subsection{Overview of RINSER} \label{ssec: overview}
Figure~\ref{fig:system} shows an overview of the RINSER workflow, from pre-processing binaries to pre-training and fine-tuning. 
It consists of three main steps: {\it Disassembly}: this step disassembles the given set of raw PE binaries and obtains assembly language-based source code from the machine-executable code. {\it API Codeprints constructor}: In this step, the disassembled source code is analyzed to construct API codeprints. An API codeprint includes (1) the API name, (2) the names and values of its input parameters, and (3) a set of assembly instructions that are semantically-related (e.g., the instruction utilizing the same registers) corresponding to each input parameter value, as highlighted in bold-font in Figure~\ref{fig:contextbuilder} (middle). \textit{Learning API contexts:} the normalized API codeprints are used to train the masked LM with BERT~\cite{devlin2018bert}, which employs a self-supervised learning technique to learn the contextual embeddings of the API codeprints. These embeddings are then used to predict the names of APIs.
The outcome of the above three steps is a pre-trained model for API prediction, depicted in Figure~\ref{fig:system} (right), that can be adjusted for diverse downstream tasks through fine-tuning.


\subsection{API codeprints construction} \label{ssec:context_builder}

RINSER automatically disassembles input binaries using IDA-Pro to extract functions and their corresponding assembly instructions. We follow three steps: (1) scan forward to locate API calls and extract API names (\S~\ref{ssec: api name and parameters extraction}); (2) backtrack from input parameter values to capture semantically relevant instructions (\S~\ref{ssec:context extraction api parameters}); and (3) convert the codeprint into a machine learning–friendly format by mapping operands such as memory and pointer addresses to symbolic tokens (e.g., \textit{mem}, \textit{ptr}) to reduce OOV issues and improve generalization (\S~\ref{ssec:symbollic mapping}). The complete process is detailed in Algorithm~\ref{algo1:context builder}.

\subsubsection{API calls and parameters extraction} \label{ssec: api name and parameters extraction}
In practice, API calls are identified by searching for the operand of the {\it call} instructions in the disassembled code~\cite{eagle2011ida,ida_flirt_tech}, e.g., {\it call RegDeleteKeyA} instruction in Figure~\ref{fig:contextbuilder} (left). 
To successfully execute a Windows API function, the necessary input parameters are supplied before calling the API (the {\it push} instructions in red-dashed rectangle in Figure~\ref{fig:contextbuilder} (left)).

The first step towards constructing API codeprints is to identify the API names and the names and values of input parameters from the disassembled code.
The input binary executable file is disassembled using IDA Pro to obtain a list of all functions (\texttt{fnList}) with respective disassembled codes  (Steps 1 - 2). To identify API names, RINSER forward scans through the disassembled code of each function and searches for {\it call} instructions, as the operand to the call instruction can potentially be the API call. If the operand to the call instruction is an external user-defined function like {\it ``call sub\_403EBC''} in Figure~\ref{fig:contextbuilder} (left), or not a {\it call} instruction, it is added to the list $F_{\text{instrs}}$, as outlined in steps 3-7 of Algorithm~\ref{algo1:context builder}.
On the other hand, if the operand is a valid Windows API call, such as {\it ``call RegDeleteKeyA''} in Figure~\ref{fig:contextbuilder} (left), the API name is extracted from the instruction and stored in the variable \texttt{api-name}, as described in steps 8-9 of Algorithm~\ref{algo1:context builder}. 

\begin{algorithm}[!ht]
\scriptsize
\caption{API codeprints construction}\label{algo1:context builder}
\begin{algorithmic}[1]
\Statex \textbf{Input: } Binary executable file $f$
\Statex \textbf{Output: }  $<\texttt{codeprint}_{1},\dots, \texttt{codeprint}_{N}>$
\Statex \textbf{Stage 1: Pre-processing}
\State $Dis(f)$ = Disassemble the input binary file using IDA-Pro
\State \texttt{fnList} = get list of all functions and respective assembly code from $Dis(f)$
\For{\texttt{fn} in \texttt{fnList}}
\State $F_{\text{instrs}}$ = []
\For{instruction \texttt{inst} in \texttt{fn}} 
\If{not a \texttt{`call'} instruction or \texttt{`call'} to an external function}
\State add \texttt{inst} to $F_{\text{instrs}}$ 
\EndIf
\If{\texttt{`call'} instruction and Windows API is called} 
\State \texttt{api-name} = get API name from \texttt{inst}
\State \texttt{params-list} = []
\State \texttt{api-context} = []
\For{\texttt{inst} in \texttt{reverse(}$F_{\text{instrs}}\texttt{)}$}\Comment{starts backtracking the source code}
\If{\texttt{`push'} instruction and annotation provided}
\State $\texttt{param}_{name}$ = parameter name from IDA-\texttt{FLIRT} 
\State $\texttt{param}_{val}$ = parameter value from \texttt{inst}
\State $\texttt{param}_{con}$ = Parameter-Value-Backtracker ($\texttt{param}_{val}$, $F_{\text{instrs}}$)
\State add $\texttt{param}_{name}$ to \texttt{params-list}
\State add $\texttt{param}_{con}$ to \texttt{api-context}
\EndIf
\EndFor
\State $\texttt{codeprint}_{n}$ = $[\texttt{api-name}, \texttt{params-list},\texttt{api-context}]$
\EndIf
\EndFor
\EndFor
\end{algorithmic}
\end{algorithm}

After extracting the API name, the next step is to extract the values and names of its input parameters. 
To achieve this, RINSER iterates through $F_{\text{instrs}}$, where the scope of $F_{\text{instrs}}$ is the assembly code of each function, in reverse order to identify instructions that pass parameters to the API function. 
As described in \S~\ref{sec : background}, parameters are pushed to the stack with a {\it push} instruction. Based on that, we use the following heuristic: if the instruction is \texttt{push}, it implies that the program is transferring parameters to the API function. 
To confirm this, we further check the annotations from IDA's FLIRT functionality~\cite{ida_flirt_tech} for the names of the parameters (as discussed in \S~\ref{sec : background}). If such annotations exist, the value and the name of the input parameter are stored in $\texttt{param}_{name}$, and $\texttt{param}_{val}$, respectively (Steps 12 - 15).
However, it is theoretically impossible to perfectly recognize API functions and parameters, so we use a heuristic-based approach to identify API input parameters. We acknowledge that this is not a perfect solution, and our heuristic-based approach is the best available solution for us, with an accuracy over 90\% (see \S~\ref{ssec:evaluation api parameters extraction}).
An API codeprint consists of the API name (obtained from step 9), its parameter names (obtained from step 15) and values (obtained from step 17), and the context for parameters (obtained from step 18).
\subsubsection{Semantic backtracking} \label{ssec:context extraction api parameters}
\begin{figure}[bt]
  \centering
  \includegraphics[width=3.4in,clip]{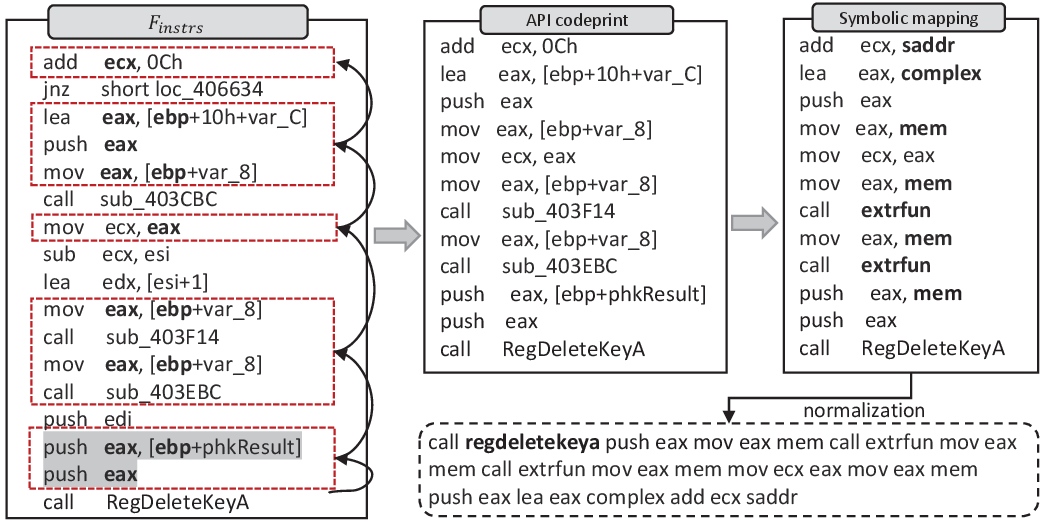}
  \caption{\small API context building from source code backtracking for RegDeleteKeyA (left). The set of semantically-related source code instructions (middle). Symbolic mapping to avoid OOV problem (right). Normalized API codeprints are shown at right-bottom.
  }\label{fig:contextbuilder}
\end{figure}

Instead of solely relying on parameter values for API prediction, e.g. in~\cite{kotov2018towards}, RINSER begins with the parameter values and traces back through $F_{\text{instrs}}$ to extract semantically-related instructions.
As a demonstration, we show how to trace back through $F_{\text{instrs}}$ and extract instructions semantically related to RegDeleteKeyA\footnote{RegDeleteKeyA takes two input parameters \textit{hKey} and \textit{lpSubKey}.} using a real-world example, which is depicted in Figure~\ref{fig:contextbuilder}.
Our intuition is if the parameter's value is stored in a CPU register, it is expected that instructions using that CPU register share the same context. Hence, RINSER identifies and extracts instructions that perform operations using common registers into groups of semantically related instructions. 
As shown in the left part of Figure~\ref{fig:contextbuilder}, the first parameter for {\it RegDeleteKeyA} is pushed onto the stack using the {\tt eax} register (highlighted in grey).  Starting with this parameter value, RINSER traces back in the code to extract semantically related instructions, as shown in Figure~\ref{fig:contextbuilder} (middle).
Algorithm~\ref{algo2:api context} outlines the backtracking process based on a parameter's value. It takes two inputs: the parameter value $\texttt{param}_{val}$ (obtained from step 16 of Algorithm~\ref{algo1:context builder}) and $F_{\text{instrs}}$. $P_{con}$ refers to semantically related instructions and $C_{regs}$ refers to CPU registers found in backtracking, both initialized steps 1-2. If $\texttt{param}_{val}$ is a CPU register (e.g., {\tt eax} in Figure~\ref{fig:contextbuilder}), it adds \texttt{inst} to $P{con}$ and its corresponding CPU register to $C_{regs}$ (steps 3 - 5). Each instruction in $F_{\text{instrs}}$ is checked. If it calls an external function (e.g., {\it ``call sub\_403EBC''} in Figure 4 (left)), it's added to $P_{con}$ (steps 7-8). External function calls are considered to provide semantics~\cite{lingmalgraph, cochard2022investigating} as they link fragmented contexts, so they are added to $P_{con}$.

If {\tt inst} is not a call instruction, its operands are examined for semantic links to other instructions via CPU register usage (steps 9–15). For instance, in Figure~\ref{fig:contextbuilder} (left), the instruction {\it ``mov eax, [ebp+phkResult]''} has two operands. The first, {\tt eax}, is already in $C_{regs}$, so the instruction is added to $P_{con}$ (steps 10–13). The second operand reveals a new register, {\tt ebp}, which is then added to $C_{regs}$ (steps 14-15), expanding the scope of the search. RINSER adaptively tracks new registers during backtracking, enabling broader context discovery. In this example, three unique registers ({\tt eax}, {\tt ebp}, {\tt ecx}) contribute to instruction correlation.

\begin{algorithm}
\scriptsize
\caption{Parameter-Value-Backtracker}\label{algo2:api context}
\begin{algorithmic}[1]
\Statex \textbf{Input: } $\texttt{param}_{val}$, $F_{\text{instrs}}$
\Statex \textbf{Output: } $P_{con}$: set of semantically-related instructions corresponding to the input parameter $\texttt{param}_{val}$
\State $P_{con}$ = []  \Comment{semantically-related instructions of $\texttt{param}_{val}$.}
\State $C_{regs}$ = [] \Comment{CPU registers involved with $\texttt{param}_{val}$.}
\If{$\texttt{param}_{val}$ is a CPU register} 
\State add \texttt{inst} to $P_{con}$
\State add $\texttt{param}_{val}$ to $C_{regs}$
\For{\texttt{inst} in \texttt{reverse(}$F_{\text{instrs}}\texttt{)}$}
\If{\texttt{`call'} instruction directed to an external function}
\State add \texttt{inst} to $P_{con}$
\EndIf
\If{not a \texttt{`call'} instruction}
\State operands = get operands from \texttt{inst}
\For{operand in operands}
\If{operand in $C_{regs}$ and \texttt{inst} not in $P_{con}$}
\State add \texttt{inst} to $P_{con}$
\EndIf
\If{operand is a CPU register and not in $C_{regs}$ and {\tt inst} in $P_{con}$}
\State add operand to $C_{regs}$
\EndIf
\EndFor
\EndIf
\EndFor
\Else 
\State add \texttt{inst} to $P_{con}$
\EndIf
\end{algorithmic}
\end{algorithm}

The outcome of Algorithm~\ref{algo2:api context} to extract semantically-related instructions $P_{con}$ for parameter $\texttt{param}_{val}$ from $F_{\text{instrs}}$. The search space for parameter semantic backtracking is limited to the set of instructions within a function, as represented by $F_{\text{instrs}}$. The API codeprint for the {\it RegDeleteKeyA} API is shown in Figure~\ref{fig:contextbuilder} (middle). The obtained API codeprints are further processed before being input into an ML model. 

\subsubsection{Post-processing API codeprints} \label{ssec:symbollic mapping}
The API codeprints from Algorithm~\ref{algo1:context builder} are further processed through symbolic mapping and normalization to replace raw assembly data with meaningful symbolic representations. This enhances pattern recognition and mitigates the out-of-vocabulary (OOV) problem, enabling better generalization during model training.
Existing approaches~\cite{ding2019asm2vec, duan2020deepbindiff, massarelli2019safe, zuo2018neural, kotov2018towards, yu2020order} apply instruction mapping and normalization to prepare data for neural networks. However, overly coarse normalization—such as removing immediate values~\cite{ding2019asm2vec, duan2020deepbindiff, zuo2018neural}—can strip away valuable context. Conversely, overly fine-grained normalization, akin to raw disassembly, often leads to an out-of-vocabulary (OOV) problem due to the explosion of unique instructions. RINSER addresses this by categorizing symbolic mappings into five key types, capturing diverse disassembly features while maintaining generalizability. The complete mapping scheme is detailed in Appendix~\ref{appendix: api symbolic mapping}.
To achieve this, memory locations are first represented using symbolic names based on the complexity of mathematical expressions used by compilers to compute the actual memory addresses of a variable. For example, as shown in Table~\ref{tab:symbolic mappings}, the expression {\it ``[esi+8]''} for computing a memory address is mapped to {\em ``mem''} because it is based on a simple addition expression, adding two memory addresses together. In contrast, a more complex expression like {\it ``[ebp+10h+var\_C]''}, which involves three operands including a register, a hexadecimal value, and a variable address, is mapped to the symbol {\it ``complex''}. Expressions with three or more operands are generally represented in this way to help the ML model better understand the relationships and patterns within the data. 

Second, disassemblers name variables based on their position relative to the saved return address~\cite{eagle2011ida}. RINSER improves on this by mapping these positions to symbolic representations based on their distance from a reference point, treating closer and farther locations differently. These positions are usually expressed in hexadecimal, and their interpretation depends on the specific value of the address.
For example, lower values like {\tt 0Ch} are mapped to smaller addresses ({\tt saddr}). Hexadecimal values with two to four characters are mapped to ({\tt maddr}), and values with more than four characters are mapped to ({\tt laddr}).
The {\it ``dword\_''} prefixes, indicating a double word (4 bytes) value, and the {\it ``off\_''} prefixes, denoting offset values, are mapped differently, as indicated in the Table~\ref{tab:symbolic mappings}.
RINSER maps user-defined functions to symbolic names (e.g., {\it ``extrfun''}).Multiple functions may be mapped to the same symbolic name. For example, user-defined function names prefixed with {\it ``sub\_40523''} are mapped to {\it ``extrfun''}.
Note that numeric values that are input as API parameters are not changed because they can provide specific context information, such as input flags or sharing options. This is in line with the approach described in~\cite{kotov2018towards}, which suggests that keeping such numeric values unchanged helps in API prediction.
Finally, we remove non-ASCII characters and punctuation and convert all instructions to lowercase. The resulting set of API codeprints, as shown in Figure 4 bottom-right, is fed to the ML model.

\subsection{Pre-training masked language model} \label{ssec: masked lm pretraining}

We propose a masked LM modeling task in which we randomly mask a certain percentage (15\%) of API context tokens and then predict the corresponding masked tokens, as shown in Figure~\ref{fig:system} right-bottom. To perform this task, we use a 12-layer Transformer with 768 hidden sizes and 12 attention heads. The intermediate size of the feed-forward networks is 3,072.
We pre-train RINSER on the training set, which contains more than 4.7 million API codeprints and is described in \S~\ref{ssec:symbollic mapping}. 
The input to the model is tokenized, and the length of each API codeprint is set to 512 tokens. 

During the pre-training process, the model runs for approximately 28,000 steps (60 epochs) with a batch size of 2,000. The learning rate is set to 2e-5, and the cosine learning rate decay technique is used. In contrast to the default BERT model, which uses a vocabulary size of 30,522, our model uses a vocabulary size of 10,522.
We chose to use a smaller vocabulary size for the following reasons:
(1) Since we are working with assembly instructions related to API calls, we believe that a smaller vocabulary size will be more effective in this case. This is because we have already applied normalization and symbolic mapping to the API codeprints, resulting in a smaller grammar; hence, we set the vocabulary size to 10,522. In general, the choice of vocabulary size is a trade-off between performance and model size, and the best choice will depend on your specific use case.
(2) A smaller vocabulary size can also make the pre-training process more efficient by reducing the amount of data that the model has to process. This can speed up the pre-training process and reduce the memory and computational resources required. Overall, the 28,000 training steps took about one day using four NVIDIA RTX A6000 GPUs with 48GB of GPU VRAM each.

\subsubsection{Fine-tuning RINSER on downstream tasks} \label{ssec:finetuning}
Fine-tuning is required to adapt a pre-trained model to specific tasks. This allows the model to adjust to the nuances and characteristics of the new task, improving accuracy and performance compared to using a pre-trained model directly.
Attempting API prediction for stripped binaries without fine-tuning will make the task challenging, resulting in poor performance due to the absence of crucial information. Our evaluation revealed that the direct use of RINSER's pre-trained model for API prediction on stripped binaries resulted in poor performance, with an accuracy of only 30\%. Therefore, fine-tuning the RINSER is necessary to enhance the performance, as it enables the model to learn the unique patterns and characteristics of stripped binaries.

\noindent\fbox{\parbox{\columnwidth}{
\underline{Normal}:\texttt{\color{blue}{ FindResourceA} \color{red} lptype \color{black} 6 \color{red} lpname \color{black} push ecx movzx ecx ax \color{red} hmodule \color{black} push edi mov edi complex push edi mov esi complex push esi}
\newline \newline
\underline{Stripped}:\texttt{\color{blue}{FindResourceA} \color{black} 6 push ecx movzx ecx ax push edi mov edi complex push edi mov esi complex push esi}
}}
\captionof{figure}{Examples of API \texttt{FindResourceA} under normal and stripped binary cases. Blue text represent API name where red-color font is API parameter names.}\label{fig:stripped and normal examples}

Normal binaries contain both the names and values of parameters, whereas stripped binaries contain only the values. 
Figure~\ref{fig:stripped and normal examples} illustrates this difference using the {\it FindResourceA} API as an example. In the normal binary case, both the API parameter names (highlighted in red) and their corresponding values are present. In contrast, the stripped binary lacks the API names (still present in IAT) and associated parameter names, leaving only the raw parameter values.
We fine-tune the RINSER model using self-supervised learning on a large corpus of stripped binaries. This pre-training followed by fine-tuning approach can also be readily applied to other API-related tasks, such as API call sequence prediction using RINSER’s pre-trained model.
  




\section{Dataset and Ground Truth Construction} \label{ssec:dataset}
\subsection{Dataset collection}
The focus of our study is to accurately predict the actual names of obfuscated APIs in malicious contexts, and we chose to use only malware binaries in our dataset for this purpose. We collected a dataset consisting of 11,098 Windows malware PE binaries. 
This is preferred because these binaries are likely to exhibit behaviors that are different from benign programs and more indicative of malicious activities. By doing so, the API call prediction model can better identify and extract APIs' malicious contexts which is important for improved performance. Additionally, the presence of benign binaries in the training data could potentially dilute the model's ability to recognize APIs used in malicious contexts. Our dataset consists solely of unpacked malware binaries, as packed malware must be unpacked to reveal its true behavior during runtime. Using only unpacked malware binaries for training and evaluation helps to ensure that our model is not biased towards packing-specific constructs in disassembly. 
The hashes for unpacked binaries were obtained from VirusShare~\cite{unpackedhashes} and were sourced from VirusTotal~\cite{virustotal}.
We used the batch mode functionality of IDA~\cite{idapro} to disassemble 11,098 raw binaries. By leveraging IDA-python~\cite{idapython} for comprehensive code analysis, we extracted 4,744,969 API codeprints from 4,214 distinct Windows APIs, comprising a total of 10,468,434 parameters (see Table~\ref{tab:dataset}). To prepare for BERT-based masked LM pre-training, we pre-processed the API codeprints using NLTK~\cite{bird2009natural} to remove special characters and punctuation marks.


\subsection{Ground truth (labelled) dataset} \label{ssec:normal_ground_truth}
The construction of the ground truth involves a two-step process. Firstly, we utilize IDA FLIRT~\cite{ida_flirt_tech} to obtain annotations for the given API codeprint. The obtained annotations do not explicitly contain the names and input parameters of an API in a structured format because FLIRT is a signature-based technology that generates annotations for disassembled code in the form of strings, symbols, API input parameter names, and more. Hence, we proceed to process the obtained annotations using our heuristics (refer to \ref{ssec: api name and parameters extraction}) to extract API names and their corresponding parameter names. The output consists of API codeprints, where each API codeprint contains the API name and a list of corresponding API parameters. 
Each parameter includes the parameter name, parameter value, and contextually-related assembly instructions, as detailed in the output of Algorithm~\ref{algo1:context builder}.
Lastly, we validated the obtained API names and corresponding parameter names obtained from FLIRT annotations against Microsoft's official documentation~\cite{microsoft_docs}, which serves as our ground truth. Microsoft documentation contain exhaustive list of all native APIs and their respective input parameters, hence serve as a ground truth. This ground truth is established by analysing over 11k binaries.

\subsubsection{Stripped binaries} Stripped binaries lack debugging symbols and often omit metadata such as parameter names, annotations, and comments. However, it still contains a readable IAT, so the API functions used by the binary are partially visible, though the internal function names are not. 
In this case, each API codeprint contains only a list of API parameters, where each parameter includes only a parameter value and corresponding set of contextually related assembly instructions. 
To create a ground truth dataset for stripped binaries, we removed both the API names and parameter names' annotations from each API codeprint (see Figure.\ref{fig:stripped and normal examples}) in the ground truth dataset obtained from normal binaries in Section \ref{ssec:normal_ground_truth}. This process resulted in API codeprints that solely contain parameter values (without parameter names) and contextually disassembled code instructions.
We test the stripped API codeprints using the fine-tuned model (outlined in \S~\ref{ssec:finetuning}), and compared against the predicted API names and the ground truth API names (derived from the removal of actual API names).

\subsubsection{Obfuscated binaries} \label{ssec:ground truth obfuscated}
The API names and corresponding parameters in obfuscated binaries are often missing and are typically represented as dummy words such as \texttt{dword\_}, \texttt{sub\_}, or register addresses, complicating the validation of RINSER's prediction results. 
We constructed an additional ground truth dataset comprising both obfuscated versions and their corresponding normal binaries. 

We employed the dynamic API resolution technique described by Suenaga et al. ~\cite{suenaga2009museum} to obfuscate APIs. Our approach crafts binaries in a manner where, rather than directly linking to Windows API functions during compilation, we choose to dynamically load the required DLLs (such as \texttt{kernel32.dll}) at runtime using functions like \texttt{LoadLibrary}, and then resolve function addresses through \texttt{GetProcAddress}.
This involved writing source code in C/C++, which was then compiled to generate the corresponding binaries. 
We obscured multiple APIs, including \texttt{CreateFile}, \texttt{WriteFile}, \texttt{CreateProcess}, and \texttt{SendMessage}, generating both obfuscated and normal versions of these APIs.
The normal version of these binaries serve as a ground truth. 
We chose to generate examples in the x86 architecture, given that the majority of malware is designed for this platform. 

We conducted the experiments on two machines:
(1) The pre-training of the BERT-based masked LM task was performed on a 64-bit Ubuntu 18.04 GPU server equipped with a 64-core Intel Xenon 4U AMD 7702 2.0GHz CPU, 32 GB of memory, 4TB of SSD storage, and 4 NVIDIA RTX A6000 48GB VRAM graphics cards,
and
(2) The evaluation of RINSER was conducted on a 64-bit Windows 10 desktop machine equipped with a 6-core CPU, 16 GB of memory, and 500GB of disk storage.


\section{Evaluation} \label{sec: evaluation}
 We aim to address four research questions (RQs):
\begin{itemize}[nolistsep,leftmargin=*]
    \item {\bf RQ 1}: How effective is RINSER in accurately extracting API codeprints from disassembly? This is crucial because inaccurate codeprints can lead to APIs being represented with incorrect input parameters and vice versa, resulting in a biased model trained on such data.
    \item {\bf RQ 2}: To what extent does RINSER succeed in accurately predicting the actual API names in normal binaries?
    \item {\bf RQ 3}: How effective is RINSER in accurately predicting the actual names of APIs for stripped binaries?
    \item {\bf RQ 4}: Can RINSER accurately predict API names in obfuscated binaries where tools like IDA fail?
\end{itemize}

\begin{table}[!t]
    \centering
    \caption{\small Dataset statistics.}
    \resizebox{2.8in}{!}{
    \begin{tabular}{lr} 
    \toprule
         \textbf{Item} & \textbf{Count} \\ \midrule
         Malware samples& 11,098 \\ 
         \#unique APIs used& 4,214 \\
         \#APIs and corresponding contexts & 4,744,969 \\
         \#number of API parameters in dataset & 10,468,434 \\
          \# of API codeprints used for training & 3,368,519 (90\%)\\
          \# of API codeprints used for testing &  336,852 (10\%) \\
        \bottomrule
    \end{tabular}
    }
\label{tab:dataset}
\end{table}

\subsection{Evaluation method and metrics}
\label{ssec: evaluation metric}
The transformer pipelines are a convenient and efficient method for using models for inference. We used the {\it fill-mask} pipeline from the Transformers library~\cite{pipelines} to perform masked language model inference. The pipeline works by masking tokens in a sequence with a special masking token and then asking the model to fill in the missing tokens.
In our case, we mask API names from API codeprints and invoke the model to predict API names.
The prediction was considered correct if the predicted API name ($\overline{w}$) matched the ground truth API name ($w$), otherwise it was considered incorrect.

We consider the following two performance metrics: (1)  API prediction accuracy: Aligned with previous research~\cite{jin2022symlm, david2020neural}, we use API prediction accuracy as our evaluation metric. It measures the proportion of correct predictions made by RINSER compared to the total number of predictions.
(2) Confidence score: We introduce a new metric, confidence score, which measures RINSER's consistency in determining the number and names of input parameters for an API (as described in \S~\ref{ssec:evaluation api parameters extraction}). The confidence score relies implicitly on the accuracy of detecting the number of input parameters and correctly identifying their names for an API.

\begin{figure}[h]
  \centering
  \includegraphics[width=3.2in,clip]{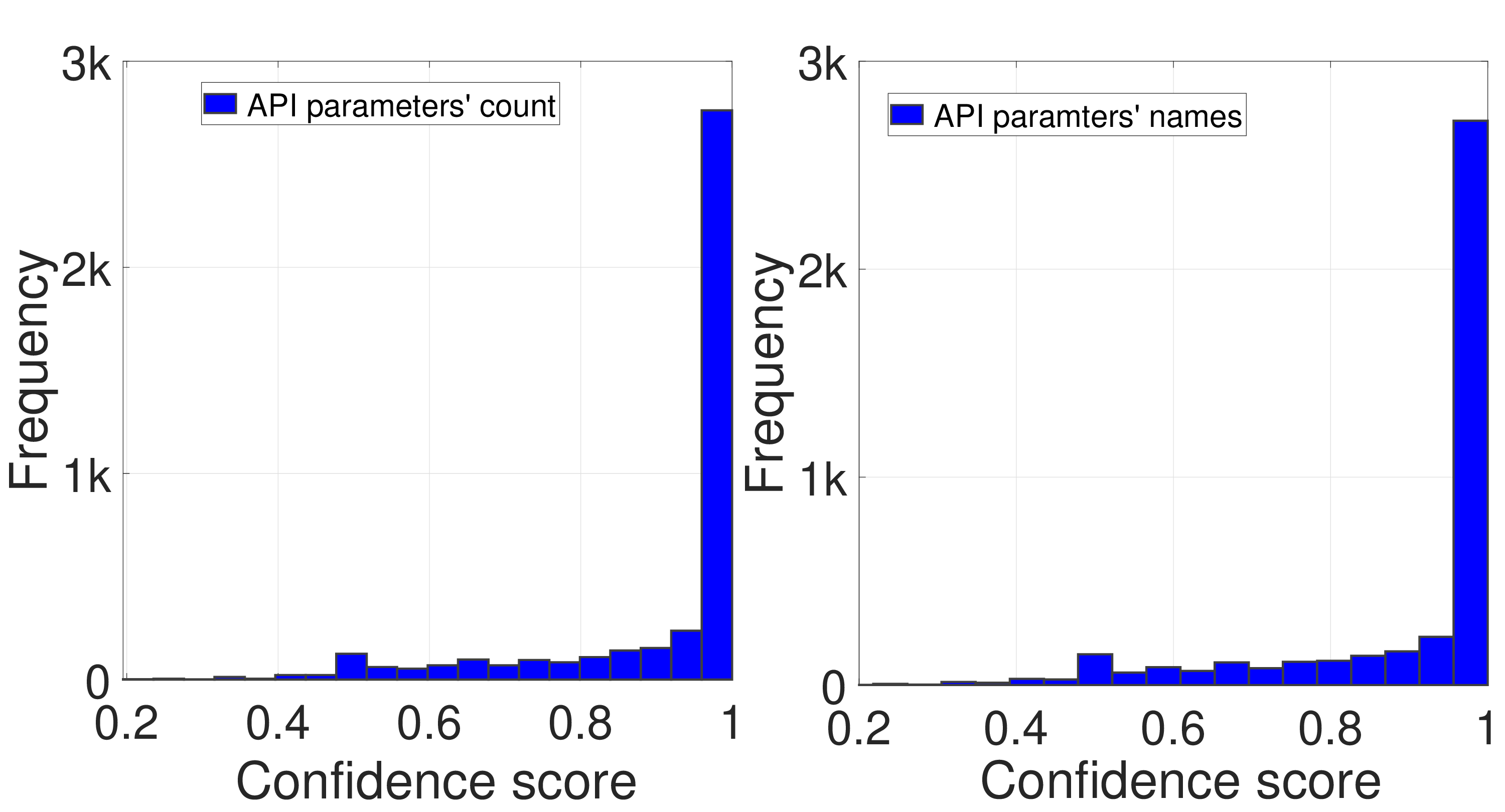}
  \caption{\small Histogram of confidence scores for APIs' parameters' count and parameters' names. 
  }\label{fig:confidence score}
\end{figure}

\subsection{API parameters extraction accuracy} \label{ssec:evaluation api parameters extraction}
We compute the confidence score of RINSER in identifying API names and their parameters by analyzing disassembly.
For example, consider the API {\it CryptReleaseContext}, which takes two input parameters: {\it hProv} and {\it dwFlags}. If this API is used 100 times in a given dataset, and RINSER correctly identifies the number (i.e., two) and names of the input parameters (i.e., {\it hProv} and {\it dwFlags}) 80 times, we can conclude that RINSER is 80\% confident that {\it CryptReleaseContext} takes two parameters, and they are named {\it hProv} and {\it dwFlags}. 


RINSER achieved an accuracy of 91.4\% in correctly identifying the number of parameters for a total of 4,123 unique Windows APIs. Additionally, RINSER achieved an accuracy of 90.5\% in correctly identifying the names of input parameters of APIs, indicating a high level of effectiveness in extracting this information from disassemblies.


Figure~\ref{fig:confidence score} presents the distribution of confidence scores for the number and names of parameters in 4,123 Windows APIs. The confidence scores, ranging from 0 to 1, are displayed on the x-axis, while the frequency or number of APIs with a given score is represented on the y-axis. 
%
%
Figure~\ref{fig:confidence score} shows that the majority of APIs have high confidence scores, as indicated by the longest bars on the right side of the graph.
The mean confidence score for the parameters count is 0.91, with a standard deviation of 0.15 and a variance of 0.02. 
Similarly, the mean confidence score for correctly identifying the names of input parameters of an API was found to be 0.90, with a standard deviation of 0.16 and a variance of 0.02.
These results were validated using Microsoft's API documentation (referenced in~\cite{microsoft_docs}). The individual information for the 4,123 APIs, including input parameter names and counts, is available with our release. 

\noindent\fbox{\parbox{\columnwidth}{
\textbf{RQ1 Answer}: Our evaluation results indicate that RINSER is highly effective at extracting the correct number (with an accuracy of 91.4\%) and names (with an accuracy of 90.5\%) of input parameters of an API from disassembled binary. 
}}
\subsection{API name prediction in normal binaries} 
\label{ssec: overall prediction small dataset}
The testing dataset, shown in Table~\ref{tab:dataset}, consists of 336,852 API codeprints. 
However, we had to filter this dataset because it included APIs that take no parameters, and RINSER does not support APIs with no input parameters (e.g., {\it GetProcessHeap}, {\it GetTickCount}) as the context information is insufficient to infer their names (as described in \S~\ref{ssec:context extraction api parameters}).
As a result, we filtered out these APIs and were left with a total of 272,334 API codeprints for evaluation of RINSER. 
We showed that RINSER correctly predicted the names of 233,598 APIs, resulting in an overall accuracy of 85.77\%. RINSER successfully predicted the actual names of 1,204 unique Windows APIs in the dataset, indicating that the model is not biased toward predicting a specific subset of APIs. 
Unlike the existing approach~\cite{kotov2018towards}, which is limited to the 25 most popular Windows APIs, RINSER is able to consider 165 times more APIs, making it a practically deployable solution for predicting API names in the case where they are obfuscated, misleading, or resolved at runtime a given input binary. Our evaluation set included APIs with input parameter counts ranging from 1 to 14, with the breakdown of APIs by parameter count 
detailed in Table~\ref{tab:parameters-breakdown} in the Appendix. Notably, approximately 84\% of the APIs in our dataset have four or fewer input parameters. Despite this diversity in the input parameter counts, RINSER's overall performance remains robust. 
RINSER's performance across different numbers of API function input parameters is detailed in Appendix~\ref{ssec:in the wild performance}.

\noindent\fbox{\parbox{\linewidth }{
\textbf{RQ2 Answer}: RINSER is highly effective in predicting the actual names of the obfuscated APIs, achieving an accuracy of 85.77\% and correctly predicting the names of 1,204 distinct Windows APIs in our dataset, which included APIs with input parameter counts ranging from 1 to 14.}}

We further evaluate RINSER's performance with respect to the number of input parameters of the APIs, and present the results in Table~\ref{tab:api prediction}. Our results show that the prediction accuracy is consistently above 86\% across all API input parameter counts. 
Additionally, we see that the prediction accuracy increases as the number of API input parameters increases (for input parameter counts ranging from 1 to 6).
Although the difference is not particularly significant, this may be because an increase in the number of input parameters also increases the amount of contextual information available, making it easier to accurately predict the API based on this context.
However, the accuracy decreases slightly to 90.55\% for APIs with more than six input parameters. This decrease, although not significant, could potentially be due to the smaller number of representative samples for these APIs in our dataset, as they may not be used as frequently.

Additionally, in Appendix~\ref{ssec: macro-average analysis}, we present a macro-average analysis to assess RINSER's performance across each API class, ensuring that our results are not skewed by an imbalance in API distribution. 
Table~\ref{tab: context-aware predictions} in Appendix~\ref{ssec: context-aware prediction} shows that embedding-based prediction, rather than exact API name matching, further improves RINSER’s performance.


\subsection{API name prediction in stripped binaries} \label{ssec:stripped apis performance}
To evaluate API prediction on stripped binaries, we derived a new dataset, see \S~\ref{ssec:in the wild performance}, by removing parameter names and debug info, as shown in Figure~\ref{fig:stripped and normal examples}. We then fine-tune our pre-trained model on this dataset for the downstream task of API prediction on stripped binaries.
After finetuning, we tested a total of 991,561 API codeprints. RINSER correctly predicted 821,895 API names, resulting in an overall prediction accuracy of 82.88\%. While we observed a degradation in the accuracy of API name prediction for stripped binaries, the performance still remained reasonable, demonstrating the effectiveness of RINSER. Table~\ref{tab:api prediction stripped binaries} summarizes our results. 

\begin{table}[!th]
    \centering
    \caption{\small API prediction against stripped binaries.}
    \resizebox{\linewidth}{!}{
    \begin{tabular}{crrrr} 
    \toprule
    \textbf{\# parameters} & \textbf{Test samples} & \textbf{Correct pred.} & \textbf{Accuracy} & \textbf{Unique APIs} \\ \midrule
    1 & 449,472 & 337,283 & 56.94\% & 319 \\
    2 & 187,084 & 164,763 & 73.34\% & 284 \\
    3 & 126,301 & 111,781 & 88.50\% & 184 \\
    4 & 103,504 & 91,120 & 88.03\% & 148 \\
    5 & 48,105 & 42,511 & 88.37\% & 83 \\
    $\geq$6 & 77,092 & 74,437 & 91.84\% & 127 \\ \midrule
    & \textbf{991,561} & \textbf{821,895} & \textbf{82.88\%} & \textbf{1,145} \\ 
    \bottomrule
    \end{tabular}
    }
\label{tab:api prediction stripped binaries}
\end{table}

We observed that the accuracy increased as the number of input parameters increased. This is likely because more input parameters provide more context, making it easier to make accurate predictions. 
For APIs taking a single input parameter, the accuracy is around 57\%, but with 6 or more input parameters, the accuracy is over 91\%. This demonstrates that having more contextual information can significantly improve the accuracy of API predictions.
An extensive ablation study in Appendix~\ref{ssec: ablation study} shows that removing context from API codeprints significantly degrades performance.

\noindent\fbox{\parbox{\linewidth}{
\textbf{RQ3 Answer}: Although RINSER performs well overall, with a prediction accuracy of 82.88\% against stripped binaries, the accuracy of its predictions varies depending on the number of input parameters. Table~\ref{tab:parameters-breakdown} provides examples of APIs and their respective input parameters.
}}

\subsection{API name prediction in obfuscated binaries} \label{ssec: obfuscation resilience}
Here, we evaluate RINSER's performance against obfuscated binaries in two scenarios: \\
\textbf{Manually crafted small dataset.} 
To evaluate, we used the ground truth dataset created in~\ref{ssec:ground truth obfuscated}. Our evaluation shows that RINSER was able to deobfuscate 66\% of the obfuscated APIs. The API codeprints extracted from disassembly were precise, leading to their accurate obfuscation. However, it's essential to highlight that these APIs were undetected by IDA, with no information present in IDA's import table and disassembly, demonstrating RINSER's effectiveness. \\
\textbf{Evaluation on a large dataset.} 
We collected a separate set of 2,918 binaries and disassembled them using IDA. We used the following criteria: (1) if the operand to the {\it call} instruction is either a {\it dword\_}, memory addresses (e.g., [{\it esi+var\_A}]) or a CPU register (e.g., {\it esi}, {\it eax}, etc.), and (2) if the input parameters are passed to the API using {\it push} or {\it mov} instructions. If these conditions were met, it was considered likely that the API was obfuscated. 
We extracted such API codeprints using the technique outlined in \S~\ref{ssec:context_builder}. This resulted in 148,685 API codeprints that did not have API names. We evaluated RINSER's pre-trained model without retraining.

Table~\ref{tab:obfuscated binary results} summarizes the dataset and evaluation results. RINSER predicted 69,921 API names which were missing in the disassembly and discovered 65 unique obfuscated APIs, which were analyzed to determine their intent (see Appendix~\ref{appendix: malware intent}). Among these 65 APIs, we found evidence of three malicious capabilities, C2 communication (e.g., {\it accept}, {\it send}, {\it gethostbyname}, {\it socket}, {\it ioctlsocket}), spying ({\it BitBlt}, {\it DrawTextExA}), and evasion ({\it Select}), by comparing them against a curated list of 442 APIs linked to nine malicious intents. Manual analysis validated RINSER’s deobfuscated API predictions.

\begin{table}[!t]
    \centering
    \caption{\small Automatic API deobfuscation results.}
    \resizebox{\linewidth}{!}{
    \begin{tabular}{c|lr} 
    \toprule
    \multirow{2}{*}{Dataset} & Total malware binaries & 2,918 \\
    & Test API codeprints (obfuscated) & 148,685 \\ \midrule
    \multirow{2}{*}{Evaluation}& Unique APIs successfully deobfuscated & 65 \\
    & Correctly predicted API names & 69,921 \\
    \bottomrule
    \end{tabular}
    }
\label{tab:obfuscated binary results}
\end{table}

\noindent\fbox{\parbox{\linewidth}{
\textbf{RQ4 Answer}: RINSER is effective in deobfuscating APIs, correctly identifying 65 unique capabilities that the commercial disassembler failed to detect.  
}}

\def\palm{\textsc{PalmTree}}
\def\trex{\textsc{Trex}}
\def\syml{\textsc{SymLM}}
\section{Related Work and Comparisons} \label{ssec: comparison and related work}

Existing approaches, such as API-Xray~\cite{cheng2021obfuscation} ocate packed malware entry points and use hardware-assisted techniques to reconstruct the IAT, a binary header structure for Windows API lookup. However, relying on specialized hardware is time-consuming, resource-intensive, and limits scalability. Similarly, Eureka~\cite{sharif2008eureka}, a generic malware analysis framework, predicts API names by tracking {\it GetProcAddress} calls and memory writes, but it struggles with obfuscation and evading dynamic analysis.
Kotov et al.~\cite{kotov2018towards} propose the most closely related work to RINSER, employing a static analysis technique known as generic API call deobfuscation (GACD). It utilizes symbolic execution and Hidden Markov Models (HMMs) to predict API names based on the parameters' values provided to the API functions. 
Following the methodology of GACD, we evaluated RINSER against the same 25 Windows APIs, comprising 25,240 test API samples from our dataset. RINSER achieved a 91.77\% API prediction accuracy, successfully predicting 23,164 API names, while GACD achieved 75.50\% accuracy. In an additional experiment, GACD employed 25 HMMs, each dedicated to a specific Windows API, achieving an overall prediction accuracy of 87.60\%. Conversely, RINSER, utilizing a single model, surpassed GACD's performance. 


Several existing approaches \palm~\cite{li2021palmtree}, \trex~\cite{pei2020trex}, and \syml~\cite{jin2022symlm} use BERT-style masked language models to generate assembly instruction embeddings for various tasks, representing the state-of-the-art.
\palm{} pretrains a general-purpose assembly language model using self-supervised learning on large unlabeled binaries, focusing on instructions and common operands while masking symbols and specific values. \trex{} builds on a BERT-style masked LM but incorporates microtraced assembly to track execution semantics and register values during pretraining, which enables fine-tuning for function similarity classification in obfuscated or optimized code. \syml{} further fine-tunes \trex{}’s pretrained model for function name prediction. Both \trex{} and \syml{} depend on source code availability, limiting applicability to malware, especially Windows PE files where source code is scarce (14\% publicly available). 



Debin~\cite{he2018debin} predicts debug information in stripped binaries using machine learning but lacks full calling context and relies on Linux-based datasets, making it unsuitable for direct comparison against malicious Windows binaries. XFL~\cite{patrick2023xfl} uses extreme multi-label learning (XML) to label function embeddings by decomposing function names into word labels, which are then reconstructed. This method requires additional overhead in dictionary creation and complex multi-label classification. XFL’s function embedding method, \textsc{Dexter}, leverages control flow graph features, making it unsuitable for direct comparison to RINSER, which focuses on API calls where such features are less relevant.

To enable a fair comparison with RINSER, we highlight key differences between our setup and prior models. Notably, \palm{} and \trex{} are trained on ELF binaries from standard Linux libraries written in C, compiled with various optimisations and architectures, and often retain headers, debug info, and symbol tables. In contrast, RINSER is trained on obfuscated, malicious PE files with such metadata typically stripped. This disparity makes direct cross-evaluation invalid.
To mitigate this, we extensive engineering work was done to re-implemented the \trex{} data pipeline and pre-trained its architecture from scratch on our dataset. However, we substituted real microtracing with dummy traces, omitting register value evolution, consistent with our understanding of \trex{}'s finetuning. \palm{} was used as-is to illustrate the performance gap resulting from dataset mismatch, with \trex{} typically outperforming it in prior work~\cite{pei2020trex}.
For evaluation, we use identical codeprints with RINSER, \trex{}, and \palm{}, extracting their embeddings and training a two-layer ReLU-activated feedforward classifier per model (adjusting only for embedding size: 768 for RINSER and \trex{}, 128 for \palm{}), over 50 epochs. We focus on the 1,024 most frequent API names, covering 96.93\% of training and 98.98\% of test data. The dataset includes 2.67M codeprints for training/validation and 1.04M for testing. As shown in Table~\ref{tab:embsbase}, RINSER outperforms both baselines, surpassing 95\% accuracy early in training, while \palm{} and \trex{} remain below 70\% even after full training.


\begin{table}[h]
    \centering
    \caption{\small Classification on top 1024 APIs after 50 epochs.}
    \resizebox{0.8\linewidth}{!}{
    \begin{tabular}{lcc} \toprule
    \textbf{Model (Emb. size)}    & \textbf{Validation Acc.} & \textbf{Testing Acc.} \\ \midrule
    \palm{} (128) & 55.75\% & 61.01\% \\
    (\textsc{PE}-)\trex{} (768) & 61.97\% & 66.48\% \\
    \textbf{RINSER (768)} & \textbf{98.67\%} & \textbf{98.99\%} \\ \bottomrule
    \end{tabular}
}
\label{tab:embsbase}
\end{table}

\section{Robustness against adversarial attacks} \label{sec: robustness}
We assess RINSER's robustness against evasion techniques proposed in~\cite{lucas2021malware,pappas2012smashing}. These methods aim to modify binaries in a way that retains their functionality while evading static analysis-based malware detectors. One such technique is Malware Makeover~\cite{lucas2021malware}, which employs optimization algorithms to iteratively generate candidate transformations that can effectively fool ML-based malware detection techniques. We evaluate RINSER's robustness against the following two types of adversarial transformations: In-place randomization (IPR) and code displacement. 

We randomly selected 1,996 malware binaries from our original dataset and applied both IPR and code displacement transformations~\cite{lucas2021malware} to construct two separate sets of transformed binaries. API codeprints were extracted from these binaries to create a test set.
Table~\ref{tab:adver results} shows the the overall performance. 
For the untransformed original binaries, RINSER achieved the prediction accuracy of 89.03\%. Additionally, it correctly identified 1,126 unique APIs, indicating RINSER's strong performance.
\begin{table}[htb]
    \centering
    \caption{\small API prediction against adversarial binaries.}
    \tabcolsep=0.05cm
    \resizebox{\linewidth}{!}{
    \begin{tabular}{llccccc} 
    \toprule
    & {\bf \# of } & \textbf{\# of unique} & \textbf{\# of} & \textbf{\# of correct} & \textbf{} & \textbf{\# correct} \\ 
    \textbf{Type} & \textbf{binaries} & \textbf{APIs} & \textbf{Codeprints} & \textbf{Pred.} & \textbf{Acc.} & \textbf{APIs} \\ \midrule
    Original & 1,996 & 2,148 & 748,135 & 667,001 & 89.03\% & 1,126 \\
    IPR-transformed & 1,343 & 1,913 & 244,466 & 215,817 & 88.28\% & 1,134 \\
    Code displacement & 1,878 & 1,898 & 237,792 & 203,151 & 85.43\% & 1,144 \\
    \bottomrule
    \end{tabular}
    }
\label{tab:adver results}
\end{table}

\subsection{IPR transformation attacks} IPR-transformed binaries employ conservative binary randomization techniques~\cite{lucas2021malware}, which include four types of transformations while preserving functionality. These transformations involve replacing instructions with equivalent ones of the same length (e.g., substituting {\it `sub eax,4'} with {\it `add eax,-4'}), reassigning registers within functions or sets of basic blocks (e.g., swapping all instances of {\it `ebx'} and {\it `ecx'}), reordering instructions based on dependencies, and modifying the order of register values pushed to and popped from the stack to maintain consistency across function calls.
These adversarial manipulations affect several aspects used by RINSER for API codeprint generation. Alterations to instructions and register changes can directly impact RINSER's semantic backtracking process for constructing API context. Similarly, instruction reordering and manipulation of {\it `push'} and {\it `pop} instructions come into play when RINSER extracts API input parameters. These transformations serve as an effective means to assess RINSER's performance against well-crafted adversarial binaries.
Our evaluation on IPR-transformed binaries achieved a high API prediction accuracy of 88.28\%, successfully predicting the names of 215,817 out of a total of 244,466 test API codeprints.

\subsection{Code displacement attacks} Code displacement~\cite{lucas2021malware} involves relocating code to a new memory section to prevent code-reuse attacks. The original code, typically at least five bytes in size, is replaced with a jump ({\it `jmp'}) instruction, redirecting program control to the relocated code. For larger displaced code, any bytes after the {\it `jmp'} are substituted with {\it `trap'} instructions to terminate code blocks. An extra {\it `jmp'} instruction is added to steer program control back to the next instruction after the displaced code, resulting in transformed control flows of binaries. 
This serves very well to evaluate RINSER's resilience against code displacement attacks using control flow manipulation. RINSER achieved 85.43\% API prediction accuracy (cf. Table~\ref{tab:adver results}) while successfully predicting 1,144 unique APIs, demonstrating that RINSER is resilient against code displacement attacks that rely on changing the control flow of a program.  
\section{Discussion and limitations} \label{sec:discussion}
\textbf{Fine-tuning RINSER.}
Pre-trained on API codeprints, RINSER can be fine-tuned for various downstream tasks. One such application is the prediction of pre-defined constant parameters, as described in \S~\ref{sec:introduction}, certain API parameters only take a small set of values. By using the context in API codeprints, RINSER can predict the values and names of these parameters. This capability could prove useful in uncovering the parameter values frequently used by malware authors to circumvent security measures.
Additionally, RINSER can also be fine-tuned based on the sequence of API calls in a binary. Appendix~\ref{ssec:qualitative evaluation} presents model explanations for the qualitative analysis.

\textbf{Limitations.}
RINSER works for Windows APIs having at-least one input parameter. Although, the fraction of such APIs is small, they still exist, e.g., {\it GetCurrentProcessId}, {\it GetTickCount}, {\it GetUserDefaultLangID}. RINSER is not able to detect such APIs. However, a plausible solution for such a problem as future work could be to fine-tune our pre-trained model with API call sequences without considering parameters. 

When Windows system libraries are statically linked (i.e., the binary contains all the required code and does not rely on external libraries) into malware, RINSER may miss API calls and parameters, as the binary avoids standard call instructions (\S~\ref{ssec: api name and parameters extraction}). However, static linking is uncommon in malware due to compatibility issues across Windows versions and increased binary size, which limits portability.


RINSER relies on {\it call} instructions to identify API calls. However, malware authors may hide this information by making calls without using the call instruction~\cite{stack_abstractinglakhotia2004}. For instance, the {\it call addr} instruction can be broken down into two {\it push} instructions and a return instruction, where the first push pushes the address of the instruction following the {\it return} instruction, and the second push pushes the address {\it addr}. In such cases, RINSER may miss key API calls.

\section{Conclusion} \label{sec:conclusion}
We have presented RINSER, a novel API deobfuscation technique based on transfer learning. It leverages a novel concept of \textit{API codeprints} to identify and extract API-specific assembly instructions from disassembly and learned in pretraining Masked Language Models to solve
API deobfuscation tasks accurately for both normal and stripped binaries.
RINSER achieved 85\% and 82\% accuracy in correctly predicting/deobfuscating API names for normal and stripped binaries, respectively.
Furthermore, RINSER is effective in accurately deobfuscating APIs correctly when commercial disassembler IDA failed to resolve API names.
Moreover, we show that RINSER is robust against instruction randomization and code displacement adversarial attacks. 

\balance
\bibliographystyle{plain}
\bibliography{bibliography}

\begin{thebibliography}{10}

\bibitem{api_hashing}
{\em API hashing}.
\newblock \url{https://malwareandstuff.com/deobfuscating-danabots-api-hashing/}.

\bibitem{idapro}
{\em hex-rays IDA Pro}.
\newblock \url{https://hex-rays.com/ida-pro/}.

\bibitem{idapython}
{\em IDAPython documentation}.
\newblock \url{https://www.hex-rays.com/products/ida/support/idapython_docs/}.

\bibitem{createfileapi}
{\em Microsoft Documentation}.
\newblock \url{https://learn.microsoft.com/en-us/windows/win32/api/fileapi/nf-fileapi-createfilew}.

\bibitem{unpackedhashes}
{\em Unpacked hashes}.
\newblock \url{https://virusshare.com/hashfiles/unpacked_hashes.md5}.

\bibitem{virustotal}
{\em ViruTotal}.
\newblock \url{https://www.virustotal.com}.

\bibitem{aghakhani2020malware}
Hojjat Aghakhani, Fabio Gritti, Francesco Mecca, Martina Lindorfer, Stefano Ortolani, Davide Balzarotti, Giovanni Vigna, and Christopher Kruegel.
\newblock When malware is packin'heat; limits of machine learning classifiers based on static analysis features.
\newblock In {\em Network and Distributed Systems Security (NDSS) Symposium 2020}, 2020.

\bibitem{alrawi2021forecasting}
Omar Alrawi, Moses Ike, Matthew Pruett, Ranjita~Pai Kasturi, Srimanta Barua, Taleb Hirani, Brennan Hill, and Brendan Saltaformaggio.
\newblock Forecasting malware capabilities from cyber attack memory images.
\newblock In {\em 30th USENIX Security Symposium (USENIX Security 21)}, pages 3523--3540, 2021.

\bibitem{bird2009natural}
Steven Bird, Ewan Klein, and Edward Loper.
\newblock {\em Natural language processing with Python: analyzing text with the natural language toolkit}.
\newblock " O'Reilly Media, Inc.", 2009.

\bibitem{theregister_malware}
Jeff Burt.
\newblock {\em How ai can help reverse engineer malware: Predicting function names of code}.
\newblock \url{https://www.theregister.com/2022/03/26/machine\_learning_malware/}.

\bibitem{cheng2018towards}
Binlin Cheng, Jiang Ming, Jianmin Fu, Guojun Peng, Ting Chen, Xiaosong Zhang, and Jean-Yves Marion.
\newblock Towards paving the way for large-scale windows malware analysis: Generic binary unpacking with orders-of-magnitude performance boost.
\newblock In {\em Proceedings of the 2018 ACM SIGSAC Conference on Computer and Communications Security}, pages 395--411, 2018.

\bibitem{cheng2021obfuscation}
Binlin Cheng, Jiang Ming, Erika~A Leal, Haotian Zhang, Jianming Fu, Guojun Peng, and Jean-Yves Marion.
\newblock $\{$Obfuscation-Resilient$\}$ executable payload extraction from packed malware.
\newblock In {\em 30th USENIX Security Symposium (USENIX Security 21)}, pages 3451--3468, 2021.

\bibitem{choi2015api}
S~Choi.
\newblock Api deobfuscator: Resolving obfuscated api functions in modern packers.
\newblock In {\em BlackHat}, 2015.

\bibitem{cochard2022investigating}
Victor Cochard, Damian Pfammatter, Chi~Thang Duong, and Mathias Humbert.
\newblock Investigating graph embedding methods for cross-platform binary code similarity detection.
\newblock In {\em 2022 IEEE 7th European Symposium on Security and Privacy (EuroS\&P)}, pages 60--73. IEEE, 2022.

\bibitem{david2020neural}
Yaniv David, Uri Alon, and Eran Yahav.
\newblock Neural reverse engineering of stripped binaries using augmented control flow graphs.
\newblock {\em Proceedings of the ACM on Programming Languages}, 4(OOPSLA):1--28, 2020.

\bibitem{devlin2018bert}
Jacob Devlin, Ming-Wei Chang, Kenton Lee, and Kristina Toutanova.
\newblock Bert: Pre-training of deep bidirectional transformers for language understanding.
\newblock {\em arXiv preprint arXiv:1810.04805}, 2018.

\bibitem{ding2019asm2vec}
Steven~HH Ding, Benjamin~CM Fung, and Philippe Charland.
\newblock Asm2vec: Boosting static representation robustness for binary clone search against code obfuscation and compiler optimization.
\newblock In {\em 2019 IEEE Symposium on Security and Privacy (SP)}, pages 472--489. IEEE, 2019.

\bibitem{dong2018understanding}
Shuaike Dong, Menghao Li, Wenrui Diao, Xiangyu Liu, Jian Liu, Zhou Li, Fenghao Xu, Kai Chen, Xiaofeng Wang, and Kehuan Zhang.
\newblock Understanding android obfuscation techniques: A large-scale investigation in the wild.
\newblock In {\em International conference on security and privacy in communication systems}, pages 172--192. Springer, 2018.

\bibitem{downing2021deepreflect}
Evan Downing, Yisroel Mirsky, Kyuhong Park, and Wenke Lee.
\newblock $\{$DeepReflect$\}$: Discovering malicious functionality through binary reconstruction.
\newblock In {\em 30th USENIX Security Symposium (USENIX Security 21)}, pages 3469--3486, 2021.

\bibitem{duan2020deepbindiff}
Yue Duan, Xuezixiang Li, Jinghan Wang, and Heng Yin.
\newblock Deepbindiff: Learning program-wide code representations for binary diffing.
\newblock In {\em Network and distributed system security symposium}, 2020.

\bibitem{eagle2011ida}
Chris Eagle.
\newblock {\em The IDA pro book, second eddition}.
\newblock no starch press, 2011.

\bibitem{capa}
Google.
\newblock {\em Mandiant's Capa}.
\newblock \url{https://github.com/fireeye/capa}.

\bibitem{he2018debin}
Jingxuan He, Pesho Ivanov, Petar Tsankov, Veselin Raychev, and Martin Vechev.
\newblock Debin: Predicting debug information in stripped binaries.
\newblock In {\em Proceedings of the 2018 ACM SIGSAC Conference on Computer and Communications Security}, pages 1667--1680, 2018.

\bibitem{ida_flirt_tech}
Hexrays.
\newblock {\em IDA FLIRT Tec.}
\newblock \url{https://hex-rays.com/products/ida/tech/flirt/in_depth/}.

\bibitem{pipelines}
{Hugging Face}.
\newblock Windows apis to intents by malware.
\newblock \url{https://malapi.io/}.
\newblock Accessed: 2022-12-19.

\bibitem{jin2022symlm}
Xin Jin, Kexin Pei, Jun~Yeon Won, and Zhiqiang Lin.
\newblock Symlm: Predicting function names in stripped binaries via context-sensitive execution-aware code embeddings.
\newblock In {\em Proceedings of the 2022 ACM SIGSAC Conference on Computer and Communications Security}, pages 1631--1645, 2022.

\bibitem{kotov2018towards}
Vadim Kotov and Michael Wojnowicz.
\newblock Towards generic deobfuscation of windows api calls.
\newblock {\em Workshop on Binary Analysis Research (BAR), NDSS}, 2018.

\bibitem{stack_abstractinglakhotia2004}
Arun Lakhotia and Eric~Uday Kumar.
\newblock Abstracting stack to detect obfuscated calls in binaries.
\newblock In {\em Source Code Analysis and Manipulation, Fourth IEEE International Workshop on}, pages 17--26. IEEE, 2004.

\bibitem{li2021palmtree}
Xuezixiang Li, Yu~Qu, and Heng Yin.
\newblock Palmtree: learning an assembly language model for instruction embedding.
\newblock In {\em Proceedings of the 2021 ACM SIGSAC Conference on Computer and Communications Security}, pages 3236--3251, 2021.

\bibitem{lingmalgraph}
Xiang Ling, Lingfei Wu, Wei Deng, Zhenqing Qu, Jiangyu Zhang, Sheng Zhang, and Tengfei Ma.
\newblock Malgraph: Hierarchical graph neural networks for robust windows malware detection.
\newblock pages 1998--2007, 2022.

\bibitem{liu2014research}
Zhenyu Liu, Yun Hu, and Lizhi Cai.
\newblock Research on software security and compatibility test for mobile application.
\newblock In {\em Fourth edition of the International Conference on the Innovative Computing Technology (INTECH 2014)}, pages 140--145. IEEE, 2014.

\bibitem{lucas2021malware}
Keane Lucas, Mahmood Sharif, Lujo Bauer, Michael~K Reiter, and Saurabh Shintre.
\newblock Malware makeover: Breaking ml-based static analysis by modifying executable bytes.
\newblock In {\em Proceedings of the 2021 ACM Asia Conference on Computer and Communications Security}, pages 744--758, 2021.

\bibitem{malapi}
MalAPI.io.
\newblock Pipelines.
\newblock \url{https://huggingface.co/docs/transformers/main_classes/pipelines}.
\newblock Accessed: 2022-12-19.

\bibitem{mandiant2022m}
Google~Cloud Mandiant.
\newblock M-trends report 2022, 2022.

\bibitem{mantovani2020prevalence}
Alessandro Mantovani, Simone Aonzo, Xabier Ugarte-Pedrero, Alessio Merlo, and Davide Balzarotti.
\newblock Prevalence and impact of low-entropy packing schemes in the malware ecosystem.
\newblock In {\em NDSS}, 2020.

\bibitem{massarelli2019safe}
Luca Massarelli, Giuseppe~Antonio Di~Luna, Fabio Petroni, Roberto Baldoni, and Leonardo Querzoni.
\newblock Safe: Self-attentive function embeddings for binary similarity.
\newblock In {\em Detection of Intrusions and Malware, and Vulnerability Assessment: 16th International Conference, DIMVA 2019, Gothenburg, Sweden, June 19--20, 2019, Proceedings 16}, pages 309--329. Springer, 2019.

\bibitem{ms_sdk}
Microsoft.
\newblock {\em sendto function (winsock.h)}.
\newblock \url{https://learn.microsoft.com/en-us/windows/win32/api/winsock/nf-winsock-sendto}.

\bibitem{microsoft_docs}
Microsoft.
\newblock {\em Technical documentation}.
\newblock \url{https://learn.microsoft.com/en-us/docs/}.

\bibitem{pappas2012smashing}
Vasilis Pappas, Michalis Polychronakis, and Angelos~D Keromytis.
\newblock Smashing the gadgets: Hindering return-oriented programming using in-place code randomization.
\newblock In {\em 2012 IEEE Symposium on Security and Privacy}, pages 601--615. IEEE, 2012.

\bibitem{park2021identifying}
Kyuhong Park, Burak Sahin, Yongheng Chen, Jisheng Zhao, Evan Downing, Hong Hu, and Wenke Lee.
\newblock Identifying behavior dispatchers for malware analysis.
\newblock In {\em Proceedings of the 2021 ACM Asia Conference on Computer and Communications Security}, pages 759--773, 2021.

\bibitem{patrick2023xfl}
James Patrick-Evans, Moritz Dannehl, and Johannes Kinder.
\newblock Xfl: Naming functions in binaries with extreme multi-label learning.
\newblock In {\em 2023 IEEE Symposium on Security and Privacy (SP)}, pages 2375--2390. IEEE, 2023.

\bibitem{pei2020trex}
Kexin Pei, Zhou Xuan, Junfeng Yang, Suman Jana, and Baishakhi Ray.
\newblock Trex: Learning execution semantics from micro-traces for binary similarity.
\newblock {\em arXiv preprint arXiv:2012.08680}, 2020.

\bibitem{pucher2022identification}
Michael Pucher.
\newblock {\em Identification of Obfuscated Function Clones in Binaries using Machine Learning}.
\newblock PhD thesis, Wien, 2022.

\bibitem{sawant2012software}
Abhijit~A Sawant, Pranit~H Bari, and PM~Chawan.
\newblock Software testing techniques and strategies.
\newblock {\em International Journal of Engineering Research and Applications (IJERA)}, 2(3):980--986, 2012.

\bibitem{sharif2008eureka}
Monirul Sharif, Vinod Yegneswaran, Hassen Saidi, Phillip Porras, and Wenke Lee.
\newblock Eureka: A framework for enabling static malware analysis.
\newblock In {\em European Symposium on Research in Computer Security}, pages 481--500. Springer, 2008.

\bibitem{suenaga2009museum}
Masaki Suenaga.
\newblock A museum of api obfuscation on win32.
\newblock {\em Symantec Security Response}, 2009.

\bibitem{vinciguerra2003experimentation}
Lori Vinciguerra, Linda Wills, Nidhi Kejriwal, Paul Martino, and Ralph Vinciguerra.
\newblock An experimentation framework for evaluating disassembly and decompilation tools for c++ and java.
\newblock In {\em 10th Working Conference on Reverse Engineering, 2003. WCRE 2003. Proceedings.}, pages 14--14. IEEE Computer Society, 2003.

\bibitem{xi2013api}
Qi~Xi, Tianyang Zhou, Qingxian Wang, and Yongjun Zeng.
\newblock An api deobfuscation method combining dynamic and static techniques.
\newblock In {\em Proceedings 2013 International Conference on Mechatronic Sciences, Electric Engineering and Computer (MEC)}, pages 2133--2138. IEEE, 2013.

\bibitem{xu2021hawkeye}
Peng Xu, Youyi Zhang, Claudia Eckert, and Apostolis Zarras.
\newblock Hawkeye: cross-platform malware detection with representation learning on graphs.
\newblock In {\em International Conference on Artificial Neural Networks}, pages 127--138. Springer, 2021.

\bibitem{yu2020order}
Zeping Yu, Rui Cao, Qiyi Tang, Sen Nie, Junzhou Huang, and Shi Wu.
\newblock Order matters: Semantic-aware neural networks for binary code similarity detection.
\newblock In {\em Proceedings of the AAAI Conference on Artificial Intelligence}, volume~34, pages 1145--1152, 2020.

\bibitem{zuo2018neural}
Fei Zuo, Xiaopeng Li, Patrick Young, Lannan Luo, Qiang Zeng, and Zhexin Zhang.
\newblock Neural machine translation inspired binary code similarity comparison beyond function pairs.
\newblock {\em arXiv preprint arXiv:1808.04706}, 2018.

\end{thebibliography}
\appendix



\section{Ablation study} \label{ssec: ablation study}
In the ablation study, we investigate the impact on the performance of RINSER in predicting API names under the following three cases:
(1) Full API codeprints: This includes parameter values, parameter names, and corresponding contextually related instructions for each input parameter.
(2) Partial API codeprints: This includes parameter names and values, but not contextually related instructions.
(3) Values-only API codeprints: This includes only parameter values.
We used the test set from \ref{ssec:in the wild performance} for this experiment. 

\begin{table}[!th]
    \centering
    \caption{\small Ablation study with and without API codeprints.}
    \resizebox{\linewidth}{!}{
    \begin{tabular}{lrrr} 
    \toprule
    \textbf{Test codeprints} & \textbf{Parameter} & \textbf{Param values} &  \textbf{Full API } \\ 
    \textbf{} & \textbf{values only} &\textbf{and names} &  \textbf{codeprints } \\ \midrule
    Model &Fine-tuned (\S~\ref{ssec:finetuning})&Original (\S\ref{ssec: masked lm pretraining})& Original (\S\ref{ssec: masked lm pretraining}) \\ \midrule
     991,561 & 19.35\% & 60.21\% & 88.80\%  \\ 
    \bottomrule
    \end{tabular}
    }
\label{tab:ablation study}
\end{table}

Table~\ref{tab:ablation study} presents RINSER's performance as the number of input parameters varies across three cases. We demonstrate that RINSER achieves an average prediction accuracy of over 88\% when full API codeprints are used. For the second case, the overall prediction accuracy is 60\%. However, performance significantly degrades for the third, reaching around 19\%, when only parameter values are used for API name prediction. Please note that, in the third case, we utilized a fine-tuned model specifically designed for stripped binaries, which means that no additional information other than parameter values is provided. However, even with the fine-tuned model, the API prediction results remain unsatisfactory, highlighting that context is crucial. without contextual information. The results above show that API codeprints with contextual instructions play an important role in accurate API name prediction.

\section{Explainability via qualitative evaluation} \label{ssec:qualitative evaluation}
To further assess  RINSER's predictions, we conducted a qualitative study of the predictions made by our pre-trained model. Specifically, we investigated the cases where RINSER made incorrect API predictions. To do this, we used our pre-trained BERT model to compute the embeddings for both the predicted and the ground truth APIs for all incorrect predictions made by our model and then calculated the cosine similarity (CS) between the predicted and ground truth API names.

By analyzing the incorrect predictions, we found that the average cosine similarity between the predicted and ground truth API names was 0.72 in cases where the predicted API names did not match the ground truth. This suggests that, even in cases where the predictions are incorrect, the predicted and ground truth APIs are still connected in context to some degree. 
Based on the types of errors made in the predictions, we grouped them into four main categories.
We present examples of incorrect predictions in each category in Table~\ref{tab:qualitative evaluation}. 
The second and third columns show the predicted and ground truth API names, respectively, and the fourth column indicates whether the input parameters of both (predicted and ground truth) APIs are identical or not. Finally, the fifth column displays the 
cosine similarity 
between the predicted and ground truth API names. 

\begin{table}[!t]
    \centering
    \caption{\small Prediction errors and cosine similarity. The predicted and ground truth API names are presented in pre-processed form.}
    \resizebox{\linewidth}{!}{
    \begin{tabular}{c|llp{2cm} >{\raggedleft\arraybackslash}p{1.3cm}} 
    \toprule
    \textbf{Errors category} & \textbf{Predicted} & \textbf{Ground truth} & \textbf{Parameters identical?} & \textbf{Cosine sim. (\%)} \\ \midrule
    \multirow{4}{*}{Encoding-specific} & loadimagea & loadimagew & Yes &0.9780 \\
    &lstrcpyw & lstrcpyna & Yes & 0.9355 \\
    &writeprivateprofilestringa & writeprivateprofilestringw & Yes & 0.9924 \\
    &gettextextentpointa & gettextextentpointw & Yes & 0.9911 \\ \midrule
    \multirow{6}{*}{Domain-related} & deletefilew & createfilew & No & 0.7659 \\ 
    &regdeletekeya & cryptdestroykey & No & 0.8899 \\
    &recv & send & Yes & 0.8523 \\
    &socket & loadimagew & No & 0.8943 \\
    &charuppera & charnexta & Yes & 0.8380 \\
    &settextcolor & setbkcolor & No & 0.9779 \\
    \midrule
    \multirow{3}{*}{Unalike} &gettickcount & drawtexta & No & 0.4816\\
    &send & callnexthookex & No & 0.4218\\
    &anglearc & scalewindowextex & No & 0.3478 \\ \midrule
    \multirow{4}{*}{Non-API} & hwnd & shcreateitemfromparsingname & NA & 0.1907 \\
    & nbar &getmenu & NA & 0.7807 \\
    &maddr &getprocaddress & NA & 0.6171 \\
    & \#\#rgn & ellipse & NA & 0.5214 \\
    \bottomrule
    \end{tabular}
    }
\label{tab:qualitative evaluation}
\end{table}

\textbf{Encoding-specific prediction errors.} 
In Microsoft Windows, some API names are distinguished by the letters `A' or `W' at the end. These letters indicate the character encoding used in the API. For example, the letter `A' after the API name signifies that ASCII character encoding is used, while the letter `W' indicates that wide character (unicode) encoding is used.
As an example, the {\it LoadImageA} API is used to load an icon, cursor, animated cursor, or bitmap using ASCII encoding~\cite{microsoft_docs}, while {\it LoadImageW} has the same functionality but with wide character encoding. Table~\ref{tab:qualitative evaluation} shows similar examples.

When the predicted and ground truth APIs differ only in character encoding, RINSER treats it as a wrong prediction according to the evaluation metric discussed in \S~\ref{ssec: evaluation metric}, even though both the predicted and ground truth APIs are actually the same, because the input parameters for these APIs and their functionalities are identical.
Category A of Table~\ref{tab:qualitative evaluation} shows examples of encoding-related mispredictions.
The cosine similarity between the predicted and ground truth API names is greater than 0.93, indicating that they are contextually related. We calculated this similarity by first obtaining the embeddings of the API names from our pre-trained model using Transformer Pipelines~\cite{pipelines}, and then calculating the cosine similarity between the resulting embedding vectors.


\textbf{Domain-related prediction errors.} 
In this case, the predicted and ground truth API names are different, meaning they may have different input parameters and functionality, but they may still be related in terms of the domain they belong to.
As shown in Table~\ref{tab:qualitative evaluation}, the predicted and ground truth APIs {\it deletefilew} and {\it createfilew} are related in terms of their domain, i.e., both the APIs are from the file system. Similarly, the {\it send} and {\it recv} APIs share the network domain.

In the graphics ({\it settextcolor}, {\it setbkcolor}, {\it charuppera}, {\it loadimagew}) and key management ({\it cryptdestroykey} and {\it regdeletekeya}) domains of the Microsoft Windows operating system, we observed a few missed predictions. The cosine similarity in this category was above average (ranging from 0.76 to 0.97), indicating that the APIs are contextually related due to their common high-level domain. However, as the APIs may be different, their input parameters may not necessarily be identical, as seen in Table~\ref{tab:qualitative evaluation}.

\textbf{Unalike prediction errors.} If the predicted and ground truth APIs have no similarities in terms of name, functionality, parameters, and domain, they are considered dissimilar. This dissimilarity is reflected in the low cosine similarity values shown in Table~\ref{tab:qualitative evaluation}. In these cases, there is no overlap between the predicted and ground truth APIs and they represent completely different concepts.

\textbf{Non-API prediction errors.} The category of non-API predictions refers to instances where the predicted APIs from our model are not actual APIs, but rather could be an input parameter (e.g., \textit{hwnd} and \textit{nbar} are parameters of graphics APIs), a mapped symbol (e.g., \textit{maddr} from symbolic mapping, shown in Table~\ref{tab:symbolic mappings}), or a tokenized word (e.g., \textit{\#\#rgn}). Since the model predictions are not valid APIs, the input parameter column is irrelevant for this category of errors. 

Building upon these findings, we propose to enhance prediction results by considering the contextual similarity of API names, as outlined in Appendix~\ref{ssec: context-aware prediction}.
\section{Macro-average analysis} \label{ssec: macro-average analysis}
\begin{figure}[h]
  \centering
  \includegraphics[width=3.2in,clip]{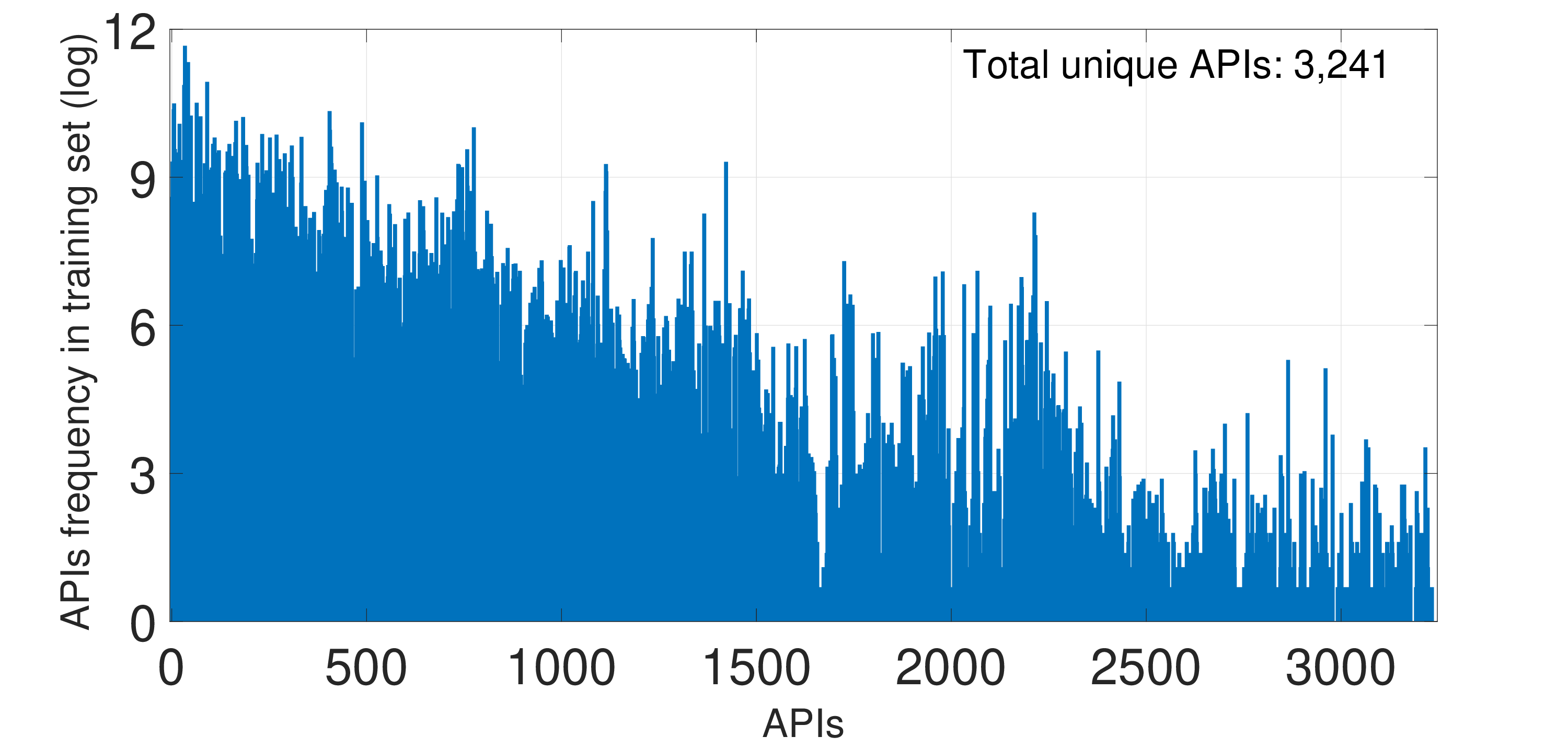}
  \caption{APIs distributions within training dataset.}\label{fig: apis training distribution}
\end{figure}

Typically, there is often an imbalance in the distribution of APIs used in malware, and this can lead to performance evaluation biases. To evaluate RINSER's performance against each API class, we conduct a macro-average analysis that computes prediction accuracy separately for each API class. 
Figure~\ref{fig: apis training distribution} illustrates the distribution (logarithmic scale) of 3,241 unique APIs within the training dataset. The x-axis in Figure~\ref{fig: apis training distribution} represents the APIs, while the y-axis denotes their respective frequencies in the training dataset. Evidently, the dataset exhibits an imbalance, with certain APIs occurring more frequently than others. The mean and standard deviation of the frequency for each API within the training set are 860 and 3,846, respectively.

\begin{figure}[h]
  \centering
  \includegraphics[width=3.2in,clip]{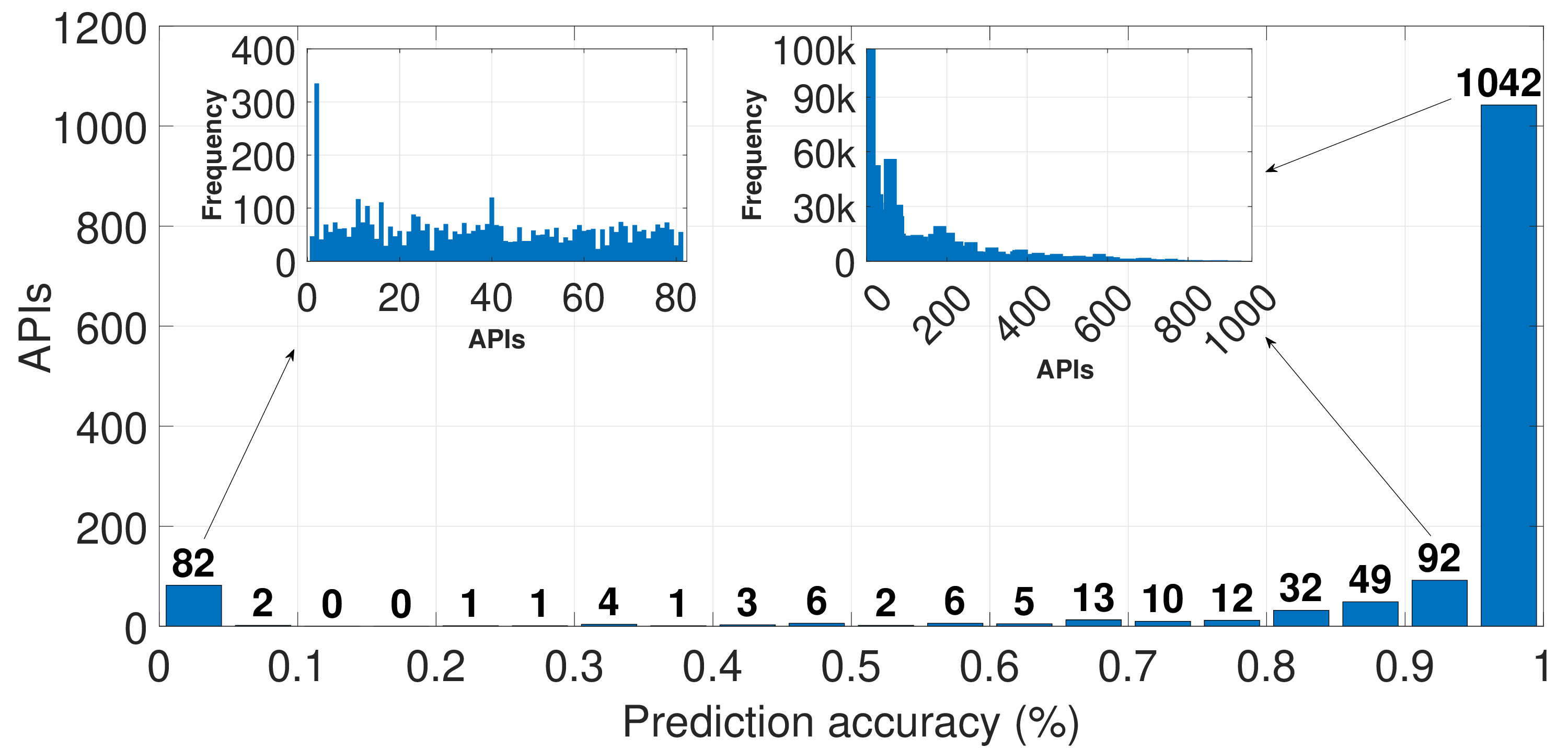}
  \caption{API prediction accuracy with an imbalanced dataset.}\label{fig: per api accuracy}
\end{figure}

To perform macro-average analysis, we used the test dataset as in \S~\ref{ssec:in the wild performance}. The test malware dataset contains a total of 1,364 unique APIs. 
Figure~\ref{fig: per api accuracy} shows the histogram of API prediction accuracy. 
Prediction accuracy is determined by the correct predictions of a specific API name by RINSER divided by the number of times that API is tested. In our test dataset, each API was observed an average of 202 times, indicating a sufficient number of test instances. We plot prediction accuracy on the x-axis and the number of unique APIs that achieved the specified prediction accuracy on the y-axis. 
In our evaluation, we observed that 1,042 unique APIs, accounting for $\approx76.3$\% of all APIs, achieved prediction accuracies of more than 95\% for correctly predicting API names. Among these, 706 APIs (51\%) achieved a perfect 100\% prediction accuracy. Conversely, 82 APIs demonstrated prediction accuracies below 1\%, underscoring the challenge RINSER faces in accurately predicting the names of these specific APIs. 

\begin{figure}[h]
  \centering
  \includegraphics[width=3.4in,clip]{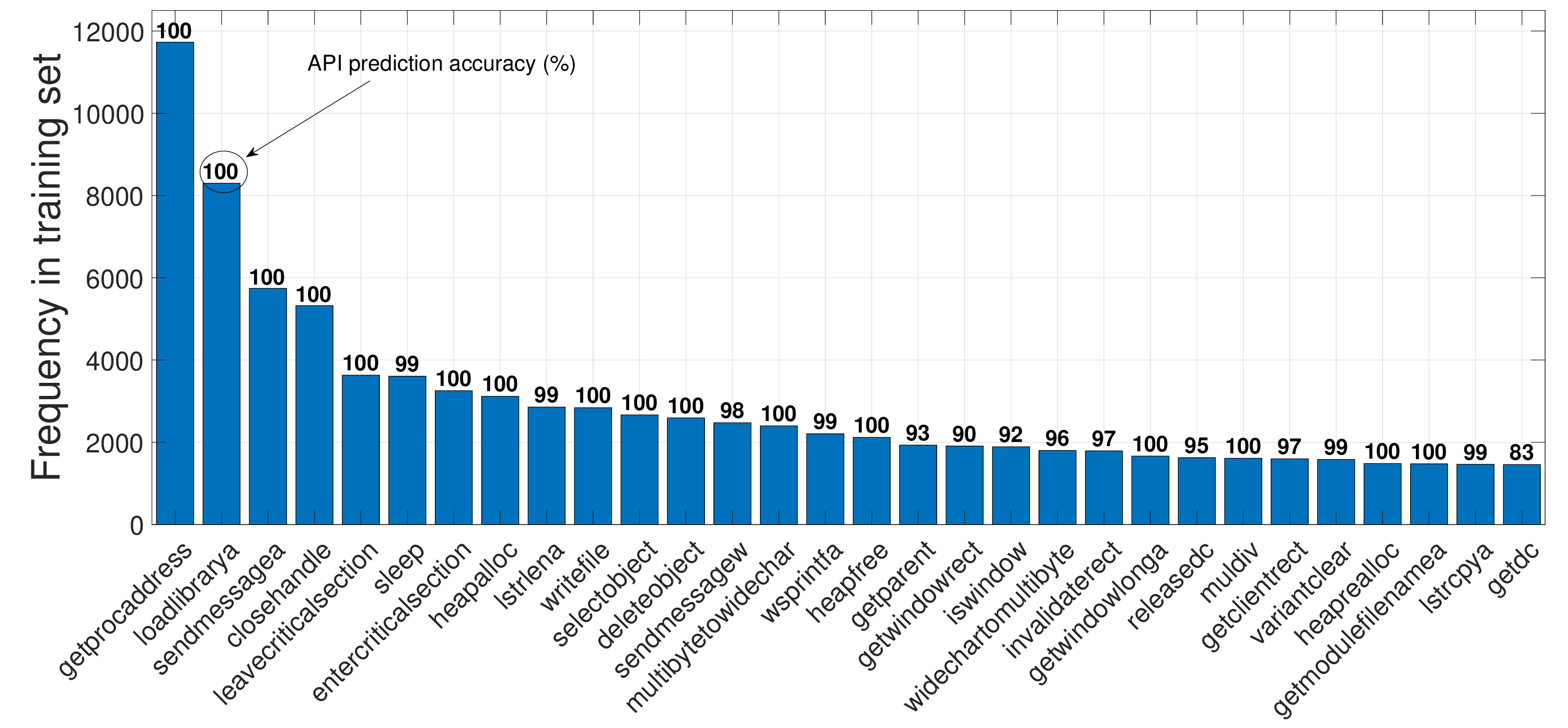}
  \caption{Top 30 most frequently used APIs in malware with respective prediction accuracies (\%).}\label{fig: top-30 apis}
\end{figure}

To delve deeper into the factors contributing to the lower prediction accuracy of the 82 APIs, we present the frequency distribution of these APIs in the training dataset, illustrated as the leftmost bar in Figure~\ref{fig: per api accuracy}. We then compare this with the frequencies of 1,134 other APIs, which are depicted by the two rightmost bars. 
We observe that the average frequency of the 82 APIs (62) in the training dataset is 35 times less than the average frequency of the 1,134 APIs (2,203), suggesting that the data imbalance within the training dataset impacts performance. Despite this imbalance in APIs within the training dataset, RINSER continues to perform well, achieving an average per-API-class prediction accuracy of 90.3\%.
The top 30 most frequent malware APIs, their frequencies in the training set, and test accuracies in the test set are detailed in Figure~\ref{fig: top-30 apis}. The most frequently used API is {\tt GetProcessAddress}, which is employed in various malicious contexts, including code injection, API hooking, import table manipulation, anti-analysis, and intercepting API calls for monitoring or modifying behavior. Similarly, malware commonly uses the {\tt LoadLibraryA} API to load DLLs for evasion purposes dynamically.

\section{Embedding-based predictions} 
\label{ssec: context-aware prediction} 
Since RINSER is based on BERT~\cite{devlin2018bert}, it may therefore be insightful to consider the model's performance when context-awareness is taken into evaluation criteria. 
More specifically, the model's prediction is considered correct if it is contextually aligned with the ground truth, even if it is not an exact match. For instance, consider the use case where we are interested in determining the capabilities of an unknown binary executable based on the APIs used in the program. Under this scenario, if the predicted API name is {\it send} whereas the ground truth is {\it recv}, it might be appropriate to consider the prediction as ``plausible'' because both APIs exhibit network capability and are contextually-related (the cosine similarity between them is more than 0.85). We have observed that many predicted APIs have a contextual relationship with the ground truth, even if the prediction is wrong, as discussed in domain-related API predictions in Table~\ref{tab:qualitative evaluation}. Similarly, encoding-specific API predictions in Table~\ref{tab:qualitative evaluation} could be considered correct predictions under any use case because the API functionalities are precisely the same but are developed with different character encodings (i.e., ASCII and Unicode).

We evaluated RINSER's context-aware prediction by relaxing the criteria for correct predictions, incorporating context into the evaluation criteria, and comparing its performance to traditional prediction standards.
To achieve this, we proposed a novel evaluation metric for context-aware predictions based on cosine similarity. 
First, we compute embedding vectors for all APIs in our dataset using Transformer's {\it feature-extraction} pipeline. 
Then, we compute pair-wise cosine similarities between APIs. Finally, we create context-based groups for APIs, i.e., for each API, we select APIs from the dataset having a cosine similarity equal to or greater than 0.91. For example, the context group of the API {\it send} contains the following network-related APIs (it is a subset of APIs): {\it connect}, {\it recv}, {\it urldownloadtofilea}, {\it internetconnectw}, {\it ftpopenfilew}.
We observed that a total of 2,537 groups were created because not all APIs will end up in groups based on our cosine similarity threshold.
\begin{table}[!th]
    \centering
    \vspace{-0.25cm}
    \caption{\small Context-aware API prediction results.}
    \resizebox{\linewidth}{!}{
    \begin{tabular}{l >{\raggedleft\arraybackslash}p{0.8in} >{\raggedleft\arraybackslash} p{0.8in}r} 
    \toprule
    \textbf{Evaluation} & \textbf{Context-aware predictions} & \textbf{Actual predictions} & \textbf{Improvement (\%)}  \\ \midrule
    Normal (\S~\ref{ssec: overall prediction small dataset}) & 88.80\% & 89.21\% & 0.41\% \\
    Stripped (\S~\ref{ssec:stripped apis performance}) & 83.67\% & 82.08\% & 1.59\%\\
    Obfuscated (\S~\ref{ssec: obfuscation resilience}) & 55.38\% & 47.03\% & 8.35\%\\
    \bottomrule
    \end{tabular}
    }
    \vspace{-0.25cm}
\label{tab: context-aware predictions}
\end{table}

At the evaluation stage, RINSER assesses the accuracy of its context-aware API predictions by comparing them with the ground truth. If the predicted API differs from the ground truth, RINSER then determines if the ground truth API is included in the context group of the predicted API. In this case, the prediction is deemed to be correct, otherwise, it is considered incorrect. As highlighted in Table~\ref{tab: context-aware predictions}, we observe that the prediction accuracy improved in all scenarios, but the greatest improvement was seen for obfuscated API calls, with an 8\% increase compared to the normal (0.41\%) and stripped (1.59\%) binary cases. This improvement can be attributed to the fact that the prediction accuracy in normal and stripped binary cases is already high, leaving little space for improvement, whereas this is not the case for obfuscated API calls.


\begin{table}[!h]
    \centering
    \caption{The breakdown of APIs based on the input parameters in our dataset. The input parameters for each API can be validated from Microsoft MSDN~\cite{microsoft_docs}.}
    \resizebox{\linewidth}{!}{
    \begin{tabular}{cr >{\raggedleft\arraybackslash}p{2.6in}} 
    \toprule
    \textbf{\# of parameters} & \textbf{\# of APIs} & \textbf{Example APIs}\\ \midrule
    1 & 812 &  \text{GetProcessId}, \text{LoadLibraryA}, \text{GetStdHandle}, \text{GetDriveTypeA}, \text{SetCurrentDirectory} \\ 
    \midrule
    2 & 667& \text{WinExec}, \text{SetFileAttributesA}, \text{GetComputerNameA}, \text{IsChild}, \text{InitializeSecurityDescriptor}, \text{CryptReleaseContext}, \text{ChangeDisplaySettingsA} \\ \midrule
    3 & 606& \text{OpenMutexA}, \text{WNetGetConnectionA}, \text{ioctlsocket}, \text{OpenInputDesktop}, \text{NetScheduleJobAdd}, \text{WriteProfileStringA} \\ \midrule
    4 & 407& \text{NetUserAdd}, \text{HttpAddRequestHeadersW}, \text{CryptImportPublicKeyInfo}, \text{CryptGenKey}, \text{NetUserGetInfo}, \text{InternetCreateUrlA} \\ \midrule
    5 & 247& \text{WriteProcessMemory}, \text{WriteProcessMemory}, \text{InternetOpenA}, \text{NtQueryObject}, \text{ryptDeriveKey}, \text{CryptCreateHash}   \\ \midrule
    6 & 138& \text{ShellExecuteA}, \text{SHSetValueA}, \text{LockFileEx}, \text{LogonUserA}, \text{WinHttpQueryHeaders}, \text{waveOutOpen}, \text{SymLoadModule} \\ \midrule
    7 & 102& \text{CreateRemoteThread}, \text{RegSetValueW}, \text{WSAConnect}, \text{WSASend}, \text{WSARecv}, \text{RegGetValueA}, \text{CreateFileW}  \\ \midrule
    8 & 58&  \text{AccessCheck}, \text{HttpOpenRequestA}, \text{acmStreamOpen}, \text{CryptDecodeObjectEx}, \text{erInstallFileA}, \text{egEnumKeyExA} \\ \midrule
    9 & 34& \text{NtWriteFile}, \text{RegCreateKeyExA}, \text{WSAIoctl}, \text{AcquireCredentialsHandleA}, \text{BitBlt}, \text{NtReadFile} \\ \midrule
    10 & 14& \text{CreateProcessA}, \text{CreateProcessW}, \text{RtlCreateUserThread}, \text{DrawStateW}, \text{EnumServicesStatusExA}, \text{ZwMapViewOfSection}  \\ \midrule
    11 & 15& \text{ChangeServiceConfigW}, \text{NtCreateFile}, \text{CreateProcessAsUserA}, \text{ZwCreateFile}, \text{CryptQueryObject}, \text{AlphaBlend} \\ \midrule
    12 & 11& \text{CreateWindowExA}, \text{RegQueryInfoKeyA}, \text{RegQueryInfoKeyW}, \text{SetDIBitsToDevice}, \text{InitializeSecurityContextA}  \\ \midrule
    13 & 6 & \text{CreateToolbarEx}, \text{CreateServiceW}, \text{CreateServiceA},  \text{DrawDibDraw}, \text{ICCompress}, \text{FCICreate} \\ \midrule
    14 & 2 & \text{CreateFontA}, \text{CreateFontW} \\
    \bottomrule
    \end{tabular}
    }
\label{tab:parameters-breakdown}
\end{table}

\section{API functions prediction analysis} 
\subsection{Symbolic mapping and normalization} 
\label{appendix: api symbolic mapping}
We show the symbolic mapping for variable values from a memory location, literal values, unresolved pointers, and functions. Numeric values remain unchanged as they exhibit real values for API parameters such as flags. Table~\ref{tab:symbolic mappings} presents the detailed symbolic transformations.

\begin{table}[!h]
    \centering
    \caption{Variable type, original code, and mapped code to normalize instructions.}
    \resizebox{0.9\linewidth}{!}{
    \begin{tabular}{l|ll} 
    \toprule
         \textbf{Input value type} & \textbf{Original values} & \textbf{Mapped to} \\ \midrule
         \multirow{ 2}{*}{Memory}&\texttt{[esi+8]}& \texttt{mem}\\
         &\texttt{[ebp+10h+var\_C]}&\texttt{complex}\\ \midrule
         \multirow{3}{*}{Hexadecimal}&\texttt{0Ch}&\texttt{saddr}\\
         &\texttt{0F6Ah}&\texttt{maddr}\\
         &\texttt{0FFFFFFF8h}&\texttt{laddr}\\ \midrule
         \multirow{3}{*}{Runtime resolution}&\texttt{unk\_} & \texttt{unknown ptr}\\
         &\texttt{\{offset, dword\_\}}& \texttt{ptr} \\
         &\texttt{off\_}& \texttt{runtime ptr} \\ \midrule
         Functions&\texttt{sub\_}& \texttt{extrfun} \\ \midrule
         Numeric&3&\texttt{3} (unchanged)\\
        \bottomrule
    \end{tabular}
    }
\label{tab:symbolic mappings}
\end{table}

\subsection{API name prediction in the wild} \label{ssec:in the wild performance}
In real-world deployment scenarios, it is common for the previously unseen binaries to be presented to the ML model. 
To assess the generalizability of RINSER, we gathered an additional set of 1,148 malware binaries and conducted disassembly using IDA~\cite{idapro}.
Following the preprocessing steps described in \S~\ref{ssec:context_builder}, we prepared a dataset of 995,422 API codeprints for evaluation.
We used the previously constructed pre-trained model of RINSER without retraining. 
RINSER correctly predicted the names of 884,018 APIs, achieving an accuracy of over 88\% over the previously unseen malware binaries. The number of distinct APIs successfully predicted in this dataset is 1,067.

\begin{table}[h]
    \centering
    \caption{\small API prediction in the wild.}
    \resizebox{\linewidth}{!}{
    \begin{tabular}{crrrr} 
    \toprule
    \textbf{\# parameters} & \textbf{Test samples} & \textbf{Correct pred.} & \textbf{Accuracy} & \textbf{Unique APIs} \\ \midrule
    1 & 449,466 & 387,035 & 86.11\% & 257 \\
    2 & 187,568 & 167,441 & 89.27\% & 244 \\
    3 & 126,297 & 116,397 & 92.16\% & 187 \\
    4 & 104,439 & 95,300 & 91.25\% & 161 \\
    5 & 48,945 & 46,576 & 95.16\% & 84 \\
    $\geq$6 & 78,707 & 71,269 & 90.55\% & 134 \\ \midrule
    & \textbf{995,422} & \textbf{884,018} & \textbf{88.80\%} & \textbf{1,067}\\ 
    \bottomrule
    \end{tabular}
    }
\label{tab:api prediction}
\end{table}


\subsection{Malware capabilities and intent} \label{appendix: malware intent}
We identified 65 obfuscated APIs during our evaluation in \ref{ssec: obfuscation resilience}. In addition, we analyze these APIs to understand the purpose of the malware and determine the functions or capabilities they provide. For example, if the malware calls APIs related to network communication or file manipulation, it could indicate that the malware is designed to communicate with other systems or manipulate files on the infected system.

We curated a list of nine malicious intents that are commonly targeted by malware and can be achieved through the use of specific APIs. The list of malicious intents, their descriptions, and the APIs associated with each intent were generated through manual effort and consultation of online resources~\cite{eagle2011ida, alrawi2021forecasting, malapi}. The full details can be found in our released repository. These intents include enumeration, injection, evasion, spying, network, anti-debugging, ransomware, dropper, and helper. We identified a total of 442 APIs that are frequently used by malicious software and categorized them into these nine intents.

\end{document}